\documentclass[twocolumn]{aastex631}

\usepackage{xcolor}
\usepackage{comment}

\newcommand{\tefftext}{$T_{\!\mbox{\scriptsize\it eff}}$}

\newcommand{\teff}{$T_{\!\mbox{\scriptsize\it eff}}$}	
\newcommand{\tefffour}{$T_{\!\mbox{\scriptsize\it eff}}^4$}	
\newcommand{\logg}{log\,$g$}
\newcommand{\loggf}{log\,$g_{\!\mbox{\tiny\it F}}$}
\newcommand{\gf}{$g_{\!\mbox{\tiny\it F}}$}

\newcommand{\mbol}{$m\,_{\!\mbox{\tiny\scriptsize bol}}$}
\newcommand{\Mbol}{$M_{bol}$}

\newcommand{\msun}{$M_\odot$}
\newcommand{\ebv}{E($B-V$)}

\newcommand{\hii}{H\,{\sc ii}\rm}

\newcommand{\nii}{[N\,{\sc ii}]}

\newcommand{\oiii}{[O\,{\sc iii}]}

\newcommand{\oii}{[O\,{\sc ii}]}
\newcommand{\oiiION}{O\,{\sc ii}}

\newcommand{\one}{\sc i\rm}

\newcommand{\opp}{O$^{++}$}

\newcommand{\cel}{{\sc cel}}
\newcommand{\rl}{{\sc rl}}
\newcommand{\eg}{{e.g.}}
\newcommand{\ie}{{i.e.}}

\newcommand{\trgb}{{\sc trgb}}
\newcommand{\bsg}{\mbox{\small BSG}}
\newcommand{\rsg}{\mbox{\small RSG}}

\newcommand{\hgamma}{H$\gamma$}

\newcommand{\hbeta}{H$\beta$}
\newcommand{\halpha}{H$\alpha$}
\newcommand{\lin}{$\,\lambda$}
\newcommand{\llin}{$\,\lambda\lambda$}

\newcommand{\rtf}{$r_{25}$}
\newcommand{\rd}{$r_d$}
\newcommand{\re}{$r_e$}

\newcommand{\eo}{$\epsilon_{\mathrm O}$}
\newcommand{\eosun}{$\epsilon_{\mathrm O,\odot}$}

\newcommand{\ohsun}{\mbox{12\,+\,log(O/H)$_\odot\,=\,$}}
\newcommand{\rtwothree}{R$_{23}$}
\newcommand{\vs}{vs.}

\renewcommand{\eg}{\mbox{e.g.}}
\renewcommand{\ie}{\mbox{i.e.}}

\newcommand{\fwhm}{\mbox{\small FWHM}}
\newcommand{\mzr}{\mbox{\small MZR}}
\newcommand{\ssc}{\mbox{\small SSC}}
\newcommand{\fglr}{\mbox{\sc fglr}}

\newcommand{\logm}{$\log(M_\star/M_\odot)$} 

\newcommand{\newhash}{%
	{\settoheight{\dimen0}{C}\kern-.05em \resizebox{!}{\dimen0}{\raisebox{\depth}{\#}}}}
\newcommand{\targetalt}[1]{\textsl{({\newhash}#1)}}		
\newcommand{\targett}[1]{\textsl{{\newhash}#1}}			
\newcommand{\target}[1]{\textsl{{\newhash}#1}}			

\newcommand{\m}{\phantom{$-$}}			

\shorttitle{NGC 2403}
\shortauthors{Bresolin et al.}

\begin{document}

\title{The metallicity and distance of NGC 2403 from blue supergiants}



\correspondingauthor{Fabio Bresolin}
\email{bresolin@ifa.hawaii.edu, kud@ifa.hawaii.edu, Miguel.Urbaneja-Perez@uibk.ac.at}

\author[0000-0002-5068-9833]{Fabio Bresolin}
\affiliation{Institute for Astronomy, University of Hawaii \\
2680 Woodlawn Drive \\
Honolulu, HI 96822, USA}

\author{Rolf-Peter Kudritzki}
\affiliation{Institute for Astronomy, University of Hawaii \\
	2680 Woodlawn Drive \\
	Honolulu, HI 96822, USA}
\affiliation{University Observatory Munich, Scheinerstr. 1, D-81679 Munich, Germany}

\author[0000-0002-9424-0501]{Miguel A. Urbaneja}
\affil{Institut f\"ur Astro- und Teilchenphysik, Universit\"at Innsbruck\\
	Technikerstr. 25/8, 6020 Innsbruck, Austria}

\begin{abstract}
We present the first quantitative spectral analysis of blue supergiant stars in the nearby galaxy NGC~2403. Out of a sample of 47 targets observed with the LRIS spectrograph at the Keck~I telescope we have extracted 16 B- and A-type supergiants for which we have data of sufficient quality to carry out a comparison with model spectra of evolved massive stars and infer the stellar parameters.
The radial metallicity gradient of NGC~2403 that we derive has a slope of
$-0.14\, (\pm 0.05)$ dex\,$r_e^{-1}$, and is in accordance with the analysis of \hii\ region oxygen abundances. We present evidence that the stellar metallicities that we obtain in extragalactic systems in general agree with the nebular abundances based on the analysis of the auroral lines, over more than one order of magnitude in metallicity.
Adopting the known relation between stellar parameters and intrinsic luminosity we find a distance modulus $\mu = 27.38 \pm 0.08$ mag. While this can be brought into agreement with Cepheid-based determinations, it is 0.14 mag short of the value measured from the tip of the red giant branch. 
We update the mass-metallicity relation secured from chemical abundance studies of stars in resolved star-forming galaxies.

\end{abstract}



\section{Introduction} \label{sec:intro}

Massive evolved stars, known as blue supergiants, can be used to trace the distribution of metals in nearby ($D$\,$<$\,10~Mpc) galaxies, by probing their surface chemical composition. The spectroscopy of individual massive stars thus affords a valuable and indispensable alternative to the emission-line analysis of the ionized gas, the long-established, standard technique used in countless studies of the near and distant universe, which unfortunately is still affected by significant and  poorly understood  systematic uncertainties (\citealt{Bresolin:2016}). The major drawback is represented by the relatively time-consuming observations needed to acquire high-quality stellar spectra, suitable for the quantitative analysis. On the other hand, the same spectroscopic data can yield information on the parent galaxy distances (\citealt{Kudritzki:2003}).

A complementary approach to obtaining present-day chemical abundances, that involves the derivation of both gas-phase and stellar values, is essential in order to untangle the difficulties in establishing the metallicity scale of star-forming galaxies, with repercussions in our understanding of the complex evolutionary processes (such as chemical mixing and galactic flows) that are at work in such systems.

Following a series of papers dealing with the quantitative analysis of stellar spectra in nearly a dozen nearby galaxies presented by our team
over the course of the past two decades  (see \citealt{Bresolin:2016}, \citealt{Kudritzki:2016}, \citealt{Urbaneja:2017}, \citealt{Berger:2018}, and references therein), covering a wide range of galactic properties (stellar mass, metallicity), we focus here on the spiral galaxy NGC~2403.

Given its small distance ($D=3.19$~Mpc, \citealt{jacobs:2009}) and the relative ease of acquiring the photometry of individual luminous stars, NGC~2403 has appeared prominently in early studies of the bright stellar content of galaxies (\citealt{Tammann:1968, Sandage:1984, Zickgraf:1991}), playing an important role in establishing the extragalactic distance scale ladder since the outset (\citealt{Hubble:1936, Tammann:1968, Sandage:1974}).

Stellar candidates in NGC~2403 were among the first extragalactic supergiant stars to be confirmed spectroscopically in a pioneering effort by \citet[see also \citealt{Humphreys:1987, Sholukhova:1998}]{Humphreys:1980}. 
More recently, \citet{Humphreys:2019} published a spectroscopic study of luminous stars in NGC~2403 and M81, aimed at the characterization of variable stars in the upper H-R diagram of these two galaxies.

In this paper we carry out the first quantitative analysis of blue supergiant (\bsg) stars in NGC~2403. Our objectives are to measure the stellar metallicities, which we use to trace the radial metallicity gradient of the system, comparing it with the result obtained from \hii\ regions, 
and to derive a spectroscopic distance to the galaxy. Our sample is
composed of 47 targets. We discuss the observations and the data reduction in Sect.~2, and characterize the stellar targets in Sect.~3, by estimating their spectral classification. In Sect.~4 we present the quantitative analysis of a subsample of 16 stars, which leads to the discussion of the galactocentric metallicity gradient and the comparison between stars and ionized gas (Sect.~5), the updated stellar mass-metallicity relation of nearby galaxies (Sect.~6) and the spectroscopic distance to NGC~2403 (Sect.~7). We conclude by summarizing our findings in Sect.~8.

\section{Observations and data reduction} \label{sec:observations}
\subsection{Target selection}
Our targets are generally fainter than the more extreme, high-luminosity objects included in the work by \citet{Humphreys:2019} and other spectroscopic surveys of luminous and variable stars in NGC~2403 (\eg\
\citealt{Humphreys:1980, Humphreys:1987}). These investigations focus on stars that are typically brighter than \mbox{$V=20$}, while approximately 80\% of our targets are fainter than this limit. The brightest blue stars in NGC~2403 are spectroscopically confirmed late-type A supergiants, appearing around \mbox{$B=18.3$} (\citealt{Zickgraf:1991}), \ie\ $M_B \simeq -9.2$. Our visually brightest spectroscopic target is also a late A supergiant star, at \mbox{$B\simeq18.4$}, and is not included in previous spectroscopic surveys of the stellar content of this galaxy.

\begin{figure*}
	\center \includegraphics[width=1\textwidth]{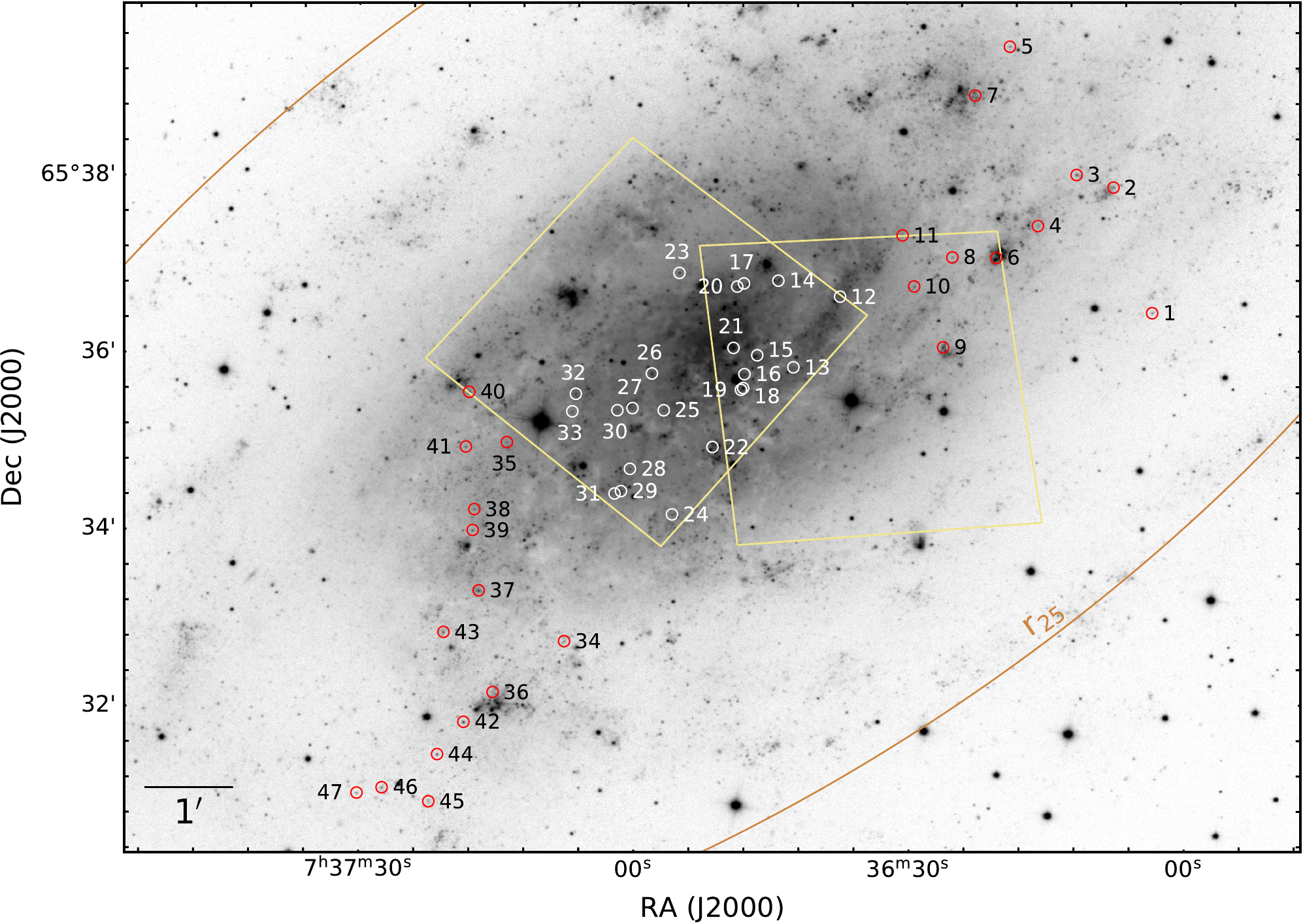}\medskip
	\caption{Position of the spectroscopic targets in a $g$-band image of NGC~2403 taken from the Sloan Digital Sky Survey. The boxes represent the footprints of the ACS observations, source of the HST photometry. The different colors of the markers and labels are only used to enhance their visibility. The outer ring represents the projected circle having a radius equal to the isophotal radius \rtf.}\label{fig:targets}
\end{figure*}

Our target selection was carried out by identifying visually bright, blue candidates from two independent sources of optical photometry. For the central region of NGC~2403 we relied on {\em Hubble Space Telescope} (HST) data obtained with the Advanced Camera for Surveys (ACS), as analyzed by the ACS Nearby Galaxy Survey Treasury (ANGST) project (\citealt{Dalcanton:2009}).
Two separate ACS Wide Field Camera pointings (see footprints in Fig.~\ref{fig:targets}) yielded magnitudes in the filters F475W, F606W, F814W (Program 10182) and F435W, F606W (Program 10579). We adopted the Johnson-Cousins $B$ and $V$ magnitudes reported by the ANGST project. These values are included in Table~\ref{table:1} (columns 7 and 8), where we summarize the positional, photometric and spectral class information of our spectroscopic targets. 

At larger galactocentric distances, not covered by the HST programs, we adopted $ugi$ photometric measurements of stellar objects we obtained from observations carried out with the MegaPrime camera at the {\em Canada-France-Hawaii Telescope} (CFHT) on Mauna Kea in September 2007 under 1-arcsec seeing conditions. 
The $g$ magnitudes and $g-i$ colors we measured are shown in Table~\ref{table:1} (columns 7 and 8), using italicized numerals, to distinguish them from the published HST photometry. 
We note that during the target selection process a few late-type stellar candidates were included in order to fill the spectroscopic multi-object masks, despite their red color indices.

Fig.~\ref{fig:targets} shows the location of the 47 targets that were selected for the spectroscopic followup, using a publicly available Sloan Digital Sky Survey (SDSS) $g$-filter image. Their celestial coordinates, measured from this image, are presented in Table~\ref{table:1}. Adopting the geometric parameters of the NGC~2403 disk 
summarized in the footnote to Table~\ref{table:1}, we calculated the galactocentric distances of the targets. In the Table these are reported normalized to both the isophotal (Column~4) and effective (Column~5) radii (\rtf\ and \re, respectively), as well as in linear units (kpc, Column~6). Our targets sit well inside the isophotal radius of NGC~2403 (partially displayed in Fig.~\ref{fig:targets}), extending radially between 0.05~\rtf\ and 0.72~\rtf.

\floattable
\begin{deluxetable}{lCCccccrcc}
\tabletypesize{\footnotesize}	
	\tablecolumns{10}
	\tablewidth{0pt}
	\tablecaption{Properties of the spectroscopic targets.\label{table:1}}
	
	\tablehead{
		\colhead{ID}	     		&
		\colhead{R.A.}	 			&
		\colhead{Decl.}	 			&
		\colhead{$r$/\rtf}			&
		\colhead{$r$/\re}			&
		\colhead{$r$}				&
		\colhead{$B$}					&
		\colhead{$B-V$}				&		
		\colhead{Spectral}			&		
		\colhead{Slit}\\[-2.5ex]
		\colhead{}	     			&
		\colhead{}	 				&
		\colhead{}	 				&
		\colhead{}					&
		\colhead{}					&
		\colhead{}					&
		\colhead{\scriptsize or}	&
		\colhead{\scriptsize or}	&		
		\colhead{type}				&		
		\colhead{}\\[-2.5ex]
		\colhead{}       			&
		\colhead{(J2000.0)}       	&
		\colhead{(J2000.0)}       	&
		\colhead{}					& 	
		\colhead{}					& 
		\colhead{(kpc)}				& 	
		\colhead{\textit{g}} 	 	& 																			
		\colhead{\textit{g$-$i}} 	& 																			
		\colhead{}			 		& 																			
		\colhead{}  } 	
	\colnumbers			
	\startdata
	\\[-4.5ex]
    01  &   07\; 36\; 03.25  &   65\; 36\; 26.5  &   0.64  &   2.38  &   6.52  & $\mathit{20.98}$  & $\mathit{-0.36}$  &   B1  &   B17 \\[-0.4ex] 
    02 $\ast$  &   07\; 36\; 07.44  &   65\; 37\; 51.8  &   0.49  &   1.80  &   4.93  & $\mathit{20.41}$  & $\mathit{-0.10}$  &   B9-A0  &   B20 \\[-0.4ex] 
    03 $\ast$  &   07\; 36\; 11.48  &   65\; 38\; 00.3  &   0.44  &   1.61  &   4.41  & $\mathit{20.21}$  & $\mathit{-0.29}$  &   B7-B8  &   B19 \\[-0.4ex] 
    04  &   07\; 36\; 15.73  &   65\; 37\; 25.8  &   0.40  &   1.47  &   4.04  & $\mathit{20.99}$  & $\mathit{-0.71}$  &   B4-B5  &   B16 \\[-0.4ex] 
    05 $\ast$  &   07\; 36\; 18.76  &   65\; 39\; 27.4  &   0.46  &   1.69  &   4.63  & $\mathit{20.86}$  & $\mathit{-0.14}$  &   B3  &   B21 \\[-0.4ex] 
    06  &   07\; 36\; 20.30  &   65\; 37\; 04.1  &   0.36  &   1.32  &   3.62  &         20.69   &         $-0.20$   &   HII region  &   B15 \\[-0.4ex] 
    07  &   07\; 36\; 22.57  &   65\; 38\; 54.4  &   0.39  &   1.43  &   3.91  & $\mathit{20.32}$  & $\mathit{ 0.81}$  &   composite  &   B18 \\[-0.4ex] 
    08 $\ast$  &   07\; 36\; 25.11  &   65\; 37\; 04.6  &   0.29  &   1.09  &   2.98  &         21.04   &         $ 0.09$   &   B0 Ib  &   B14 \\[-0.4ex] 
    09  &   07\; 36\; 26.14  &   65\; 36\; 03.7  &   0.36  &   1.34  &   3.69  &         19.86   &         $-0.07$   &   B1  &   B11 \\[-0.4ex] 
    10 $\ast$  &   07\; 36\; 29.30  &   65\; 36\; 45.0  &   0.26  &   0.95  &   2.60  &         20.60   &         $ 0.10$   &   B9  &   B12 \\[-0.4ex] 
    11  &   07\; 36\; 30.56  &   65\; 37\; 19.6  &   0.23  &   0.84  &   2.31  & $\mathit{20.96}$  & $\mathit{-0.23}$  &   B2  &   B13 \\[-0.4ex] 
    12  &   07\; 36\; 37.41  &   65\; 36\; 38.1  &   0.16  &   0.58  &   1.59  &         21.40   &         $ 0.16$   &   composite  &   B10 \\[-0.4ex] 
    13  &   07\; 36\; 42.52  &   65\; 35\; 50.2  &   0.16  &   0.61  &   1.67  &         21.52   &         $ 0.21$   &   composite  &   B07 \\[-0.4ex] 
    14  &   07\; 36\; 44.16  &   65\; 36\; 49.0  &   0.09  &   0.35  &   0.95  &         20.17   &         $ 0.01$   &   B2-B3  &   B09 \\[-0.4ex] 
    15  &   07\; 36\; 46.47  &   65\; 35\; 58.4  &   0.09  &   0.34  &   0.93  &         21.36   &         $-0.07$   &   B1 Ib  &   B06 \\[-0.4ex] 
    16 $\ast$  &   07\; 36\; 47.88  &   65\; 35\; 45.6  &   0.11  &   0.39  &   1.08  &         20.15   &         $ 0.24$   &   B4  &   A28 \\[-0.4ex] 
    17  &   07\; 36\; 47.93  &   65\; 36\; 47.1  &   0.09  &   0.32  &   0.89  &         21.55   &         $ 0.10$   &   indeterminate  &   B08 \\[-0.4ex] 
    18 $\ast$  &   07\; 36\; 48.02  &   65\; 35\; 36.2  &   0.13  &   0.48  &   1.33  & $\mathit{18.38}$  & $\mathit{ 0.21}$  &   A7-F0  &   A27 \\[-0.4ex] 
    19  &   07\; 36\; 48.26  &   65\; 35\; 35.0  &   0.13  &   0.49  &   1.33  &         20.27   &         $ 0.14$   &   B7+WC  &   B05 \\[-0.4ex] 
    20 $\ast$  &   07\; 36\; 48.67  &   65\; 36\; 45.0  &   0.09  &   0.31  &   0.86  &         20.07   &         $ 0.18$   &   B8-B9  &   A31 \\[-0.4ex] 
    21  &   07\; 36\; 49.08  &   65\; 36\; 03.4  &   0.05  &   0.17  &   0.46  &         21.27   &         $ 0.16$   &   B3 Ib  &   A29 \\[-0.4ex] 
    22  &   07\; 36\; 51.38  &   65\; 34\; 56.2  &   0.21  &   0.78  &   2.15  &         20.50   &         $ 0.31$   &   B4  &   B04 \\[-0.4ex] 
    23 $\ast$  &   07\; 36\; 54.99  &   65\; 36\; 54.2  &   0.17  &   0.62  &   1.69  &         20.19   &         $-0.01$   &   B2  &   A30 \\[-0.4ex] 
    24  &   07\; 36\; 55.80  &   65\; 34\; 10.5  &   0.31  &   1.14  &   3.14  &         20.63   &         $ 0.19$   &   B+WNL  &   A20 \\[-0.4ex] 
    25 $\ast$  &   07\; 36\; 56.70  &   65\; 35\; 21.1  &   0.11  &   0.40  &   1.11  &         20.13   &         $ 0.23$   &   A3  &   A24 \\[-0.4ex] 
    26  &   07\; 36\; 58.00  &   65\; 35\; 46.0  &   0.07  &   0.27  &   0.73  &         20.70   &         $-0.11$   &   O8-O9  &   A25 \\[-0.4ex] 
    27  &   07\; 37\; 00.12  &   65\; 35\; 22.6  &   0.11  &   0.41  &   1.13  &         20.80   &         $-0.05$   &   B0  &   A23 \\[-0.4ex] 
    28 $\ast$  &   07\; 37\; 00.40  &   65\; 34\; 41.4  &   0.20  &   0.74  &   2.04  &         20.55   &         $ 0.16$   &   A2  &   B02 \\[-0.4ex] 
    29 $\ast$  &   07\; 37\; 01.38  &   65\; 34\; 26.2  &   0.24  &   0.88  &   2.41  &         20.03   &         $ 0.13$   &   B8  &   A19 \\[-0.4ex] 
    30  &   07\; 37\; 01.77  &   65\; 35\; 21.1  &   0.12  &   0.45  &   1.24  &         21.02   &         $ 0.13$   &   B9-A0 Ib  &   B03 \\[-0.4ex] 
    31  &   07\; 37\; 02.08  &   65\; 34\; 24.6  &   0.24  &   0.89  &   2.43  &         20.76   &         $-0.02$   &   B1-B2  &   B01 \\[-0.4ex] 
    32  &   07\; 37\; 06.32  &   65\; 35\; 32.2  &   0.16  &   0.61  &   1.66  &         21.14   &         $ 0.28$   &   A1-A2  &   A22 \\[-0.4ex] 
    33  &   07\; 37\; 06.70  &   65\; 35\; 20.3  &   0.17  &   0.61  &   1.69  &         20.83   &         $-0.02$   &   B0-B1  &   A21 \\[-0.4ex] 
    34  &   07\; 37\; 07.57  &   65\; 32\; 44.5  &   0.49  &   1.79  &   4.92  & $\mathit{20.91}$  & $\mathit{-0.54}$  &   B1-B2 Ib  &   A12 \\[-0.4ex] 
    35  &   07\; 37\; 13.86  &   65\; 34\; 59.4  &   0.24  &   0.90  &   2.47  & $\mathit{20.94}$  & $\mathit{-0.37}$  &   indeterminate  &   A17 \\[-0.4ex] 
    36  &   07\; 37\; 15.36  &   65\; 32\; 09.9  &   0.55  &   2.03  &   5.57  & $\mathit{20.92}$  & $\mathit{ 1.22}$  &   indeterminate  &   A08 \\[-0.4ex] 
    37  &   07\; 37\; 16.91  &   65\; 33\; 18.8  &   0.39  &   1.43  &   3.92  & $\mathit{19.57}$  & $\mathit{-0.31}$  &   B3+WN  &   A11 \\[-0.4ex] 
    38 $\ast$  &   07\; 37\; 17.40  &   65\; 34\; 13.9  &   0.30  &   1.11  &   3.06  & $\mathit{20.86}$  & $\mathit{-0.30}$  &   A1-A2  &   A14 \\[-0.4ex] 
    39  &   07\; 37\; 17.57  &   65\; 33\; 59.8  &   0.32  &   1.18  &   3.24  & $\mathit{20.73}$  & $\mathit{ 0.26}$  &   B7-B8  &   A13 \\[-0.4ex] 
    40  &   07\; 37\; 17.98  &   65\; 35\; 33.5  &   0.32  &   1.17  &   3.21  &         20.53   &         $ 0.05$   &   B1 Ib  &   A18 \\[-0.4ex] 
    41 $\ast$  &   07\; 37\; 18.34  &   65\; 34\; 56.4  &   0.29  &   1.09  &   2.98  & $\mathit{20.67}$  & $\mathit{-0.23}$  &   B7  &   A15 \\[-0.4ex] 
    42  &   07\; 37\; 18.55  &   65\; 31\; 49.7  &   0.59  &   2.19  &   6.01  & $\mathit{20.18}$  & $\mathit{-0.17}$  &   B0  &   A06 \\[-0.4ex] 
    43 $\ast$  &   07\; 37\; 20.74  &   65\; 32\; 50.6  &   0.45  &   1.66  &   4.57  & $\mathit{20.01}$  & $\mathit{-0.35}$  &   B1  &   A09 \\[-0.4ex] 
    44  &   07\; 37\; 21.41  &   65\; 31\; 27.7  &   0.64  &   2.37  &   6.51  & $\mathit{20.39}$  & $\mathit{ 0.00}$  &   A0 III  &   A04 \\[-0.4ex] 
    45  &   07\; 37\; 22.36  &   65\; 30\; 55.8  &   0.72  &   2.66  &   7.31  & $\mathit{20.91}$  & $\mathit{ 0.65}$  &   indeterminate  &   A03 \\[-0.4ex] 
    46  &   07\; 37\; 27.46  &   65\; 31\; 05.2  &   0.69  &   2.54  &   6.97  & $\mathit{20.64}$  & $\mathit{-0.04}$  &   B3 Ib  &   A02 \\[-0.4ex] 
    47 $\ast$  &   07\; 37\; 30.19  &   65\; 31\; 01.5  &   0.70  &   2.57  &   7.04  & $\mathit{20.90}$  & $\mathit{ 0.09}$  &   B8-B9 Ib  &   A01 \\[-0.4ex] 
    \\[-2.5ex]
	\enddata
	\tablecomments{Normalized galactocentric distances adopt the following disk geometry: i\,=\,63~deg, PA\,=\,124~deg (\citealt{de-Blok:2008}), \rtf\,=\,657 arcsec (\citealt{Kendall:2011}) and \re\,=\,178 arcsec (\citealt{Rogers:2021}). Stars identified with the $\ast$ symbol are those analyzed in Sect.~\ref{sec:quantitative}. Unless the luminosity class is specified, the spectra are consistent with the Ia class. Columns 7 and 8 report either the HST (non italicized) or the CFHT (italicized) photometry.}
\end{deluxetable}

\subsection{Spectroscopy}
Spectra of the 47 targets were acquired with the Low Resolution Imaging Spectrometer (LRIS, \citealt{Oke:1995}) at the Keck~I telescope on 2010 Dec 31 -- 2011 Jan 02. While both blue and red channel multi-object data were gathered, our paper only presents results based on the blue channel data, obtained with the 600/4000 grism, and covering the approximate wavelength range 3300--5600~\AA. We took advantage of an identical setup earlier in 2010 for the spectroscopy of \bsg s in M81 (\citealt{Kudritzki:2012}), a galaxy placed at a comparable distance to NGC~2403.

Two separate masks (A and B) were cut with 1.2 arcsec-wide slits, yielding a spectral resolution of $\sim$5~\AA. The seeing conditions varied considerably throughout the course of the observing run, from 0.8 to 1.8 arcsec. In the course of the data analysis we decided to utilize only the images with the best image quality, around 1 arcsec \fwhm, which limited the total effective integration time to 3\,h (mask A) and 3.5\,h (mask B), respectively.

The data reduction, carried out with {\sc iraf}\footnote{{\sc iraf} is distributed by the National Optical Astronomy Observatories, which are operated by the Association of Universities for Research in Astronomy, Inc., under cooperative agreement with the National Science Foundation.}, included the standard procedures of bias subtraction, flat field correction and wavelength calibration. Individual frames were registered and combined. The spectral extractions were finally normalized to unity for the subsequent steps: spectral classification and, when feasible, quantitative analysis. For convenience, Table~\ref{table:1} (which is ordered by increasing RA) retains in its last column  the original nomenclature of the extracted spectra, which indicates the parent mask (A or B) and the running number of the slit.

At the wavelength of \hgamma\ (4340~\AA) the signal-to-noise ratio of the final spectra peaks around S/N\,=\,70  for the brightest target \targetalt{18}\footnote{For the remainder of the paper we will indicate our targets using the ID number in column 1 of Table~\ref{table:1}, prepended by the hash symbol.} at $B=18.4$, decreasing to S/N\,$\simeq$\,40 at $B=20$ and S/N\,$\simeq$\,20 at $B=21$. 

\section{Spectral classification} \label{sec:classification}
For the MK classification of our targets we relied on the monograph by \citet{Gray:2009} and the associated digital spectral standard library (see \citealt{Gray:2014}). Column 9 of Table~\ref{table:1} summarizes our results, and shows that for several targets we could not successfully assign a spectral classification. We describe these cases as either `composite', where we can identify specific lines from multiple types, or `indeterminate', where such identification could not be carried out, possibly as a result of an inferior signal-to-noise ratio.
For targets that are considered to be bona fide stars the classification is often provided in terms of a range of types in Table~\ref{table:1}, reflecting the intrinsic uncertainty in unequivocally assigning a spectral class (\ie\ due to weak lines and poor signal-to-noise ratio) and the independent assessment by two of the authors.

Nebular emission (either from localized \hii\ regions or diffuse ionized gas) is present in virtually all cases (the \oii\lin3727 emission line is detected in all spectra), and heavy contamination of the low-order stellar Balmer lines (\hbeta, \hgamma) is therefore common. One target \targetalt{06} is an \hii\ region whose spectrum does not display stellar lines.

\begin{figure*}
	\center \includegraphics[width=1\textwidth]{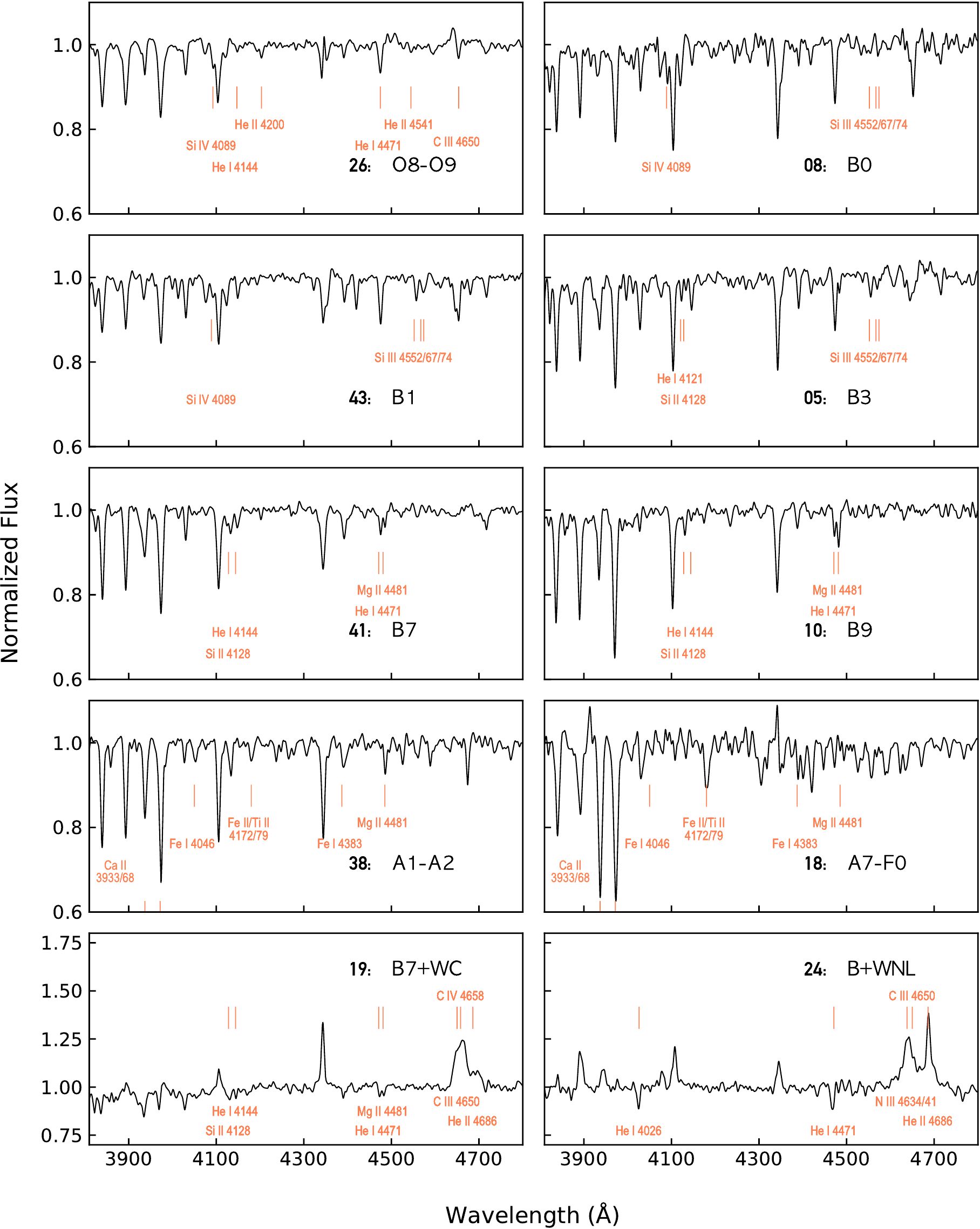}\medskip
	\caption{Examples of the stellar classification analysis. Each subpanel includes both the identification from Table~\ref{table:1} (column 1) and the spectral type. The spectra are smoothed to a \fwhm\ resolution of 5\,\AA.
	The main spectral features used for the classification are identified.}\label{fig:spectra}
\end{figure*}

The comparison with the spectral standards, in particular the information contained in the line widths, yielded a Ia luminosity classification for the majority of the targets. The exceptions are represented by eight objects of class Ib and a single (foreground) class III star \targetalt{44}. 

Three emission line stars (two type N and one type C Wolf-Rayet stars) are included in our target list -- in all cases we can discern absorption lines arising from the presence of a companion.
The spectral class exhibited by the remaining bona fide stars ranges from O8-O9 \targetalt{26} to A7-F0 \targetalt{18}: the photometric, color-based selection procedure we adopted proved quite effective in providing a sample of about 35 early-type, BA-type supergiants in NGC~2403 (more than 70\% of the full spectroscopic sample). In Fig.~\ref{fig:spectra} we display a selection of spectra spanning the full range of spectral classes, and include the identification of the features used during the classification procedure.

We have two objects in common with \citet{Humphreys:2019}: their targets 10579-x1-3 (corresponding to our target \targett{20}) and 10182-pr-16 (our target \targett{29}). In the case of the former \citet{Humphreys:2019} indicate the presence of nebular emission superposed on top of stellar absorption, but do not provide an estimate of the  spectral class. We classify \targett{20} as a B8-B9 supergiant.
We confirm the B8 supergiant classification assigned by \citet{Humphreys:2019} to the second star in common.
Both stars are included in the quantitative analysis that follows.

\section{Quantitative analysis} \label{sec:quantitative}
For 16 of our targets (identified by the asterisk symbols in column 1 of Table~\ref{table:1}) we were able to carry out a quantitative spectral analysis, aimed at deriving stellar parameters (\logg\ and \teff) and metallicities. The stellar metallicities, measured for supergiant stars in NGC~2403 for the first time, are compared in Sect.~\ref{sec:comparison} with the gas-phase metallicity of \hii\ regions in the same galaxy. In addition, the stellar parameters are employed in Sect.~\ref{sec:distance} to obtain the distance to NGC~2403.

The technique we adopt to analyze the spectra of extragalactic \bsg s has been covered in detail in previous papers (\eg\ \citealt{Urbaneja:2005, Kudritzki:2008, Kudritzki:2012, Kudritzki:2016, Hosek:2014}). Here we provide a concise description of our method, which relies on comparing the observed, normalized stellar spectra to synthetic spectra of \bsg s. 
Different grids of stellar models and line formation calculations are used for early B-type (B0--B4 for the stars in Table~\ref{table:1}) and A-type (B7 and later) stars. Following \citet{Urbaneja:2017}, we refer to these two groups as OB and BA supergiants, respectively. \\

\noindent
\textit{OB supergiants} -- The model spectra of the early-type stars are calculated with the {\sc fastwind} code (\citealt{Puls:2005}), which accounts for the presence of stellar winds in the expanding atmospheres of these hot objects. 
We calculated a grid of non-LTE, line blanketed atmosphere models covering a large parameter space, namely effective temperatures (\tefftext) from 15,000 to 30,000~K,
surface gravities (\logg) corresponding to evolutionary stages extending from the main sequence to the Eddington limit,
and surface chemical compositions where the metallicity [Z] ranges from $-1.0$ to 0.3
dex. The line formation is also calculated by {\sc fastwind}. 
\\[-5pt]

\noindent
\textit{BA supergiants} -- In this case the grid of hydrostatic, line-blanketed model atmospheres described by \citet{Kudritzki:2008, Kudritzki:2012} is adopted, covering wide ranges in \tefftext\ (7900--16,000~K), \logg\ (0.8--3.0~dex, cgs) and metallicity [$Z$] ($-$1.3 to 0.5 dex). We scale the abundance of the metals to the solar ratios, adopting the solar composition by \citet{Grevesse:1998}, except for oxygen, for which we refer to \citet{Allende-Prieto:2001}.
The synthetic spectra are calculated in non-LTE, following \citet{Przybilla:2006}. 
\\[-5pt]

\begin{figure*}
	\epsscale{1.15}\plottwo{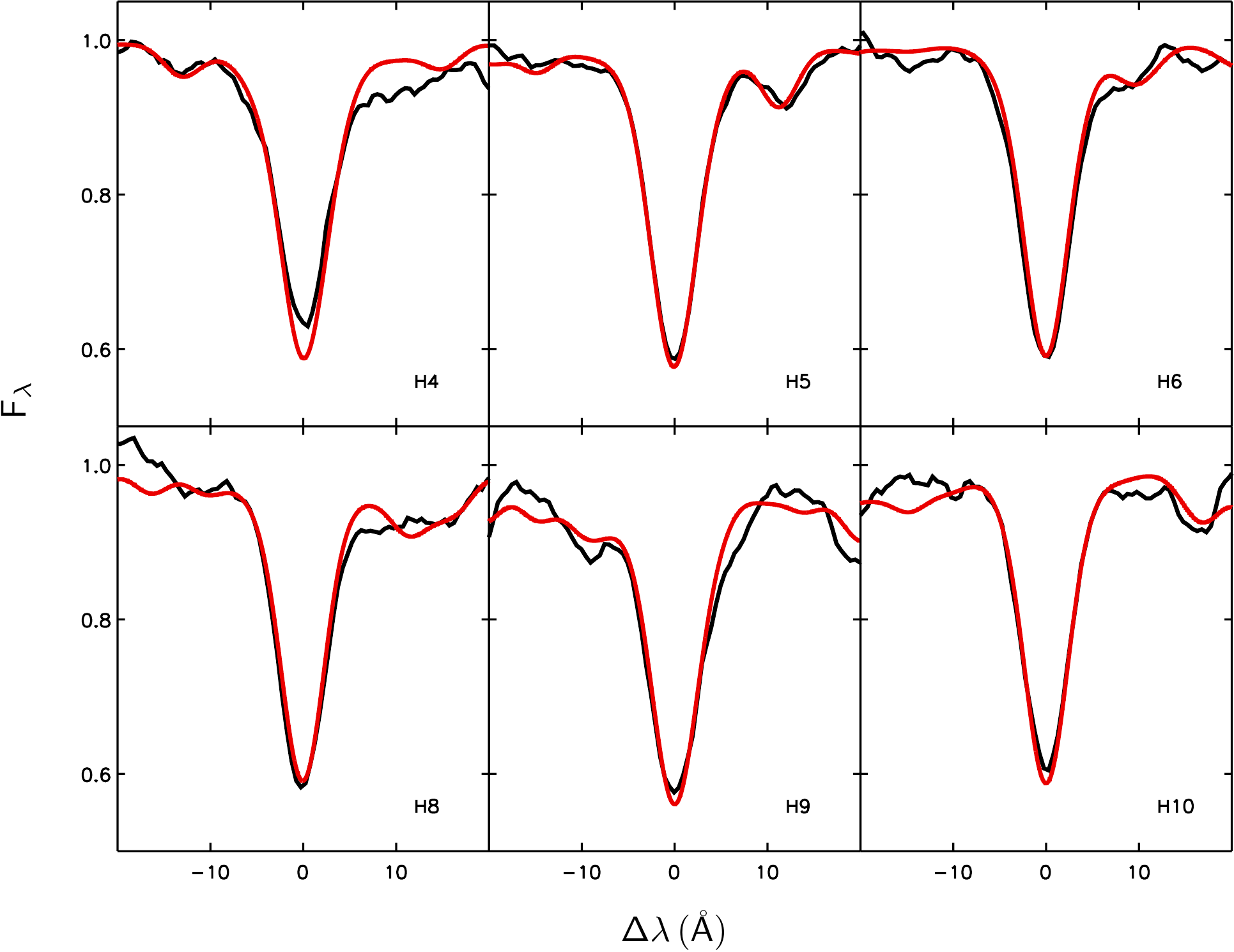}{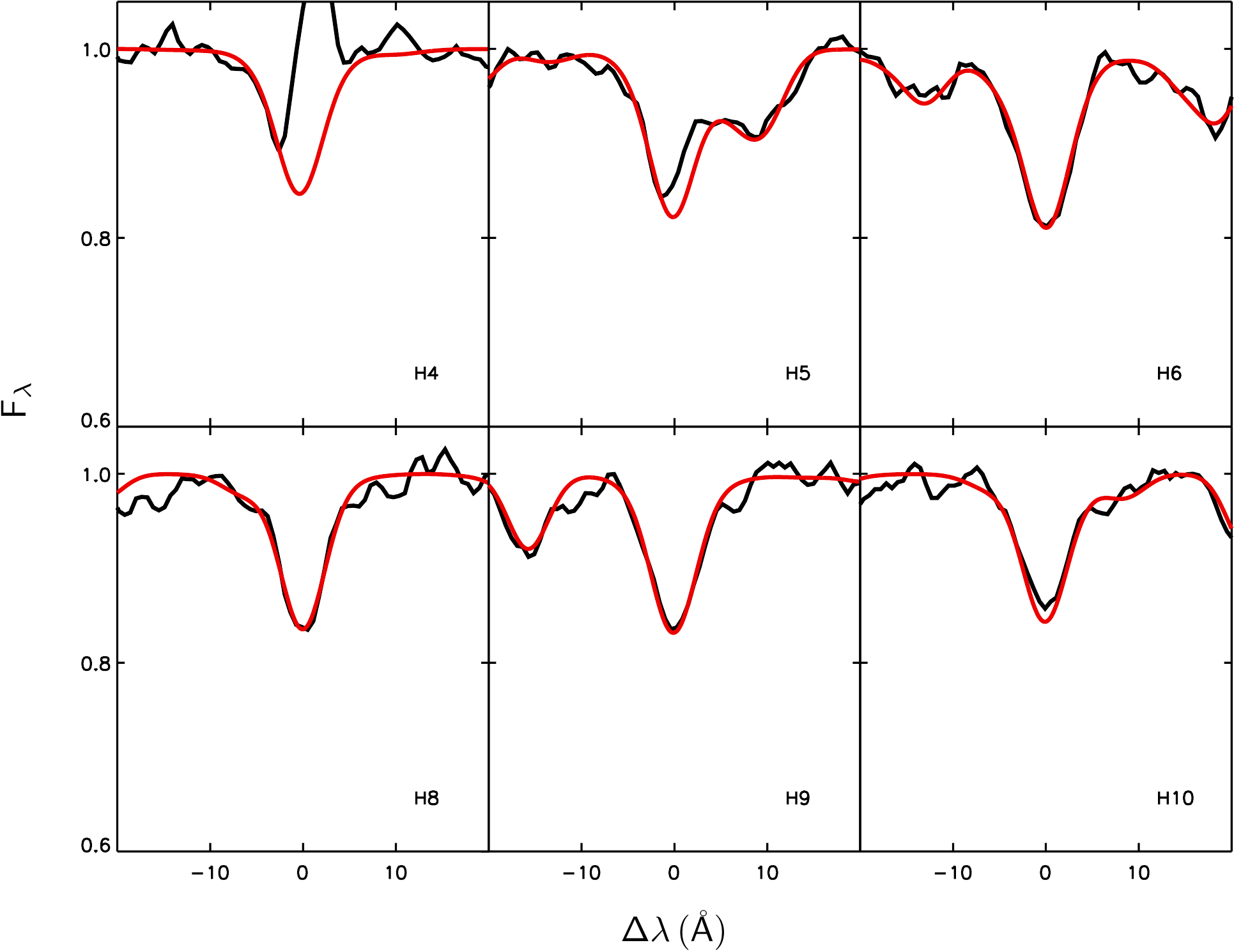}\medskip
	\caption{Fits of the model (red curve) to the observed (black curve) Balmer lines H4 to H10 for the A2~Ia star \target{28} (left) and the B1~Ia star \target{43} (right). The abscissa reports the distance in \AA\ from the line center. Some moderate nebular line  contamination can be noticed in the case of the \hbeta\ (H4) line for \target{28}, while for \target{43} the contamination is much stronger, affecting also \hgamma\ (H5).}\label{fig:BalmerFit}
\end{figure*}

A $\chi^2$ minimization technique is used inside strategically selected wavelength ranges across the supergiant spectra in order to find the best-fitting values of \tefftext, \logg\ and [$Z$]. The Balmer lines (H4 to H10 are usable, given the observed wavelength range and spectral resolution)
provide the surface gravity diagnostic, once the stellar temperature value is established. As stated earlier, these lines are often contaminated by ionized gas emission, but thanks to the decreasing equivalent width of the emission component with increasing order of the line, we can generally make satisfactory model fits to \hgamma\ (H5) and higher-order lines. An example of Balmer line fitting is shown in Fig.~\ref{fig:BalmerFit}.

\begin{figure}
	\center \epsscale{1.15}\plotone{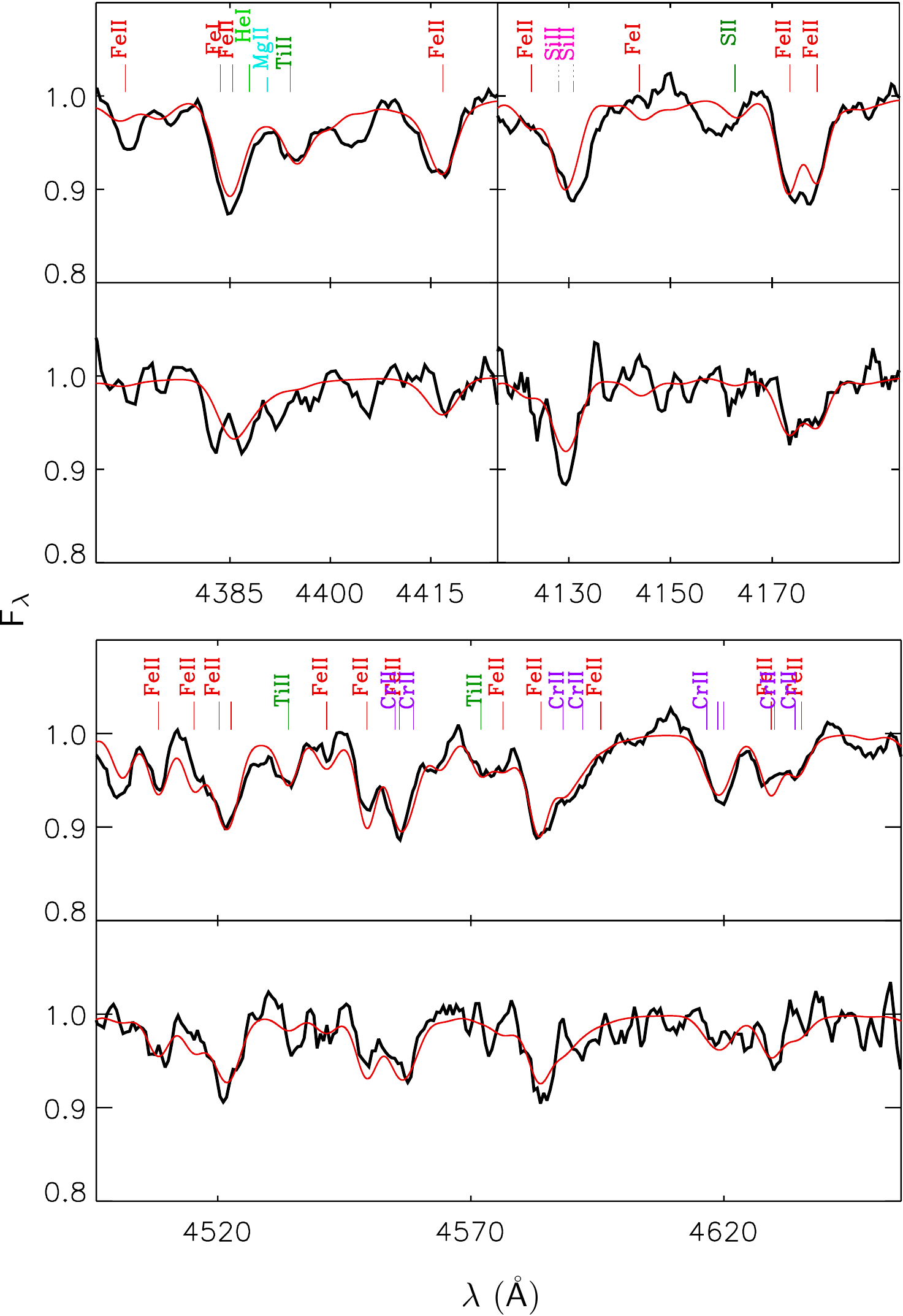}\medskip
	\caption{Fits of the model (red curve) to the observed (black curve) metal lines of stars \target{28} (type A2~Ia; at the top in each panel) and \target{38} (A1-A2~Ia; bottom) inside three wavelength windows. Lines from the main atomic species used to calculate the synthetic spectra are identified.}\label{fig:MetalFit}
\end{figure}

Due to the relatively low spectral resolution and limited signal-to-noise ratio of our data, in order to measure the stellar metallicity we do not rely on individual spectral lines, but rather on the simultaneous
fit of various spectral features that are due to metals, as shown in Fig.~\ref{fig:MetalFit}. 
It should be noted that in the case of the OB stars the metal lines found in the observed spectral range refer mostly to C, N and the $\alpha$ elements O, Mg and Si. For the BA stars the elements responsible for most of the metal lines are the $\alpha$ elements Mg, Si, Ti and the iron peak elements Cr and Fe. The overlap in atomic species represented in the spectra of OB and BA supergiants increases our confidence that the metallicities we assign to our targets are on the same absolute scale between the two groups, despite the differences in generating the models.

In order to complete the characterization of our targets, the
interstellar reddening value for each star is evaluated by comparing
the broad-band colors at our disposal ($B-V$, $u-g$, $g-i$) to the
spectral energy distribution of the best-fitting model. We adopt the
\citet{Cardelli:1989} extinction law with a total-to-selective
absorption coefficient $R_V$\,=\,$A_V/$\ebv\,=\,3.2 and $R_g$\,=\,$A_g/$E($u-g$)\,=\,3.89 .
The best-fitting solution also provides the bolometric correction (BC), used to calculate the apparent bolometric magnitude \mbol.

In the case of the targets with CFHT photometry, we make use of the information from both color indices available to us. The adopted reddening is the average E*($u-g$) = 0.5\,E($u-g$) + 0.6\,E($g-i$), where the coefficient for the second term proceeds from simulations involving our model spectra and the Cardelli reddening law.
We only use either E($u-g$) or E($g-i$) in those cases where the photometry yields negative E($g-i$) 
(\target{41} and \target{05}) or E($u-g$) (\target{18}) values, respectively.

Some of our targets are located inside \hii\ regions and, as a consequence, their photometry could be affected by line emission. However, in order to avoid severe nebular contamination of the stellar spectra, we have restricted our analysis to objects where this impact is minimal. For the most extreme cases we use the strengths of the emission lines relative to the stellar continuum in conjunction with the filter functions and estimate the potential effects on the broad-band photometry to be smaller than 0.01 mag.

Table~\ref{table:2} summarizes the stellar parameters we have calculated for the 16 targets. The value of the flux-weighted gravity \loggf\,=\,\logg $\,-\, 4\log$(\teff/10$^4$\,K) in column 4 is used in Sect.~\ref{sec:distance} for the derivation of the spectroscopic distance to NGC~2403. The errors quoted in the Table are obtained from the $\chi^2$ minimization technique (\citealt{Hosek:2014}) and, in the case of \logg, the sensitivity of the model Balmer lines to variations in surface gravity. 

\begin{figure}
	\epsscale{1.15}\center\plotone{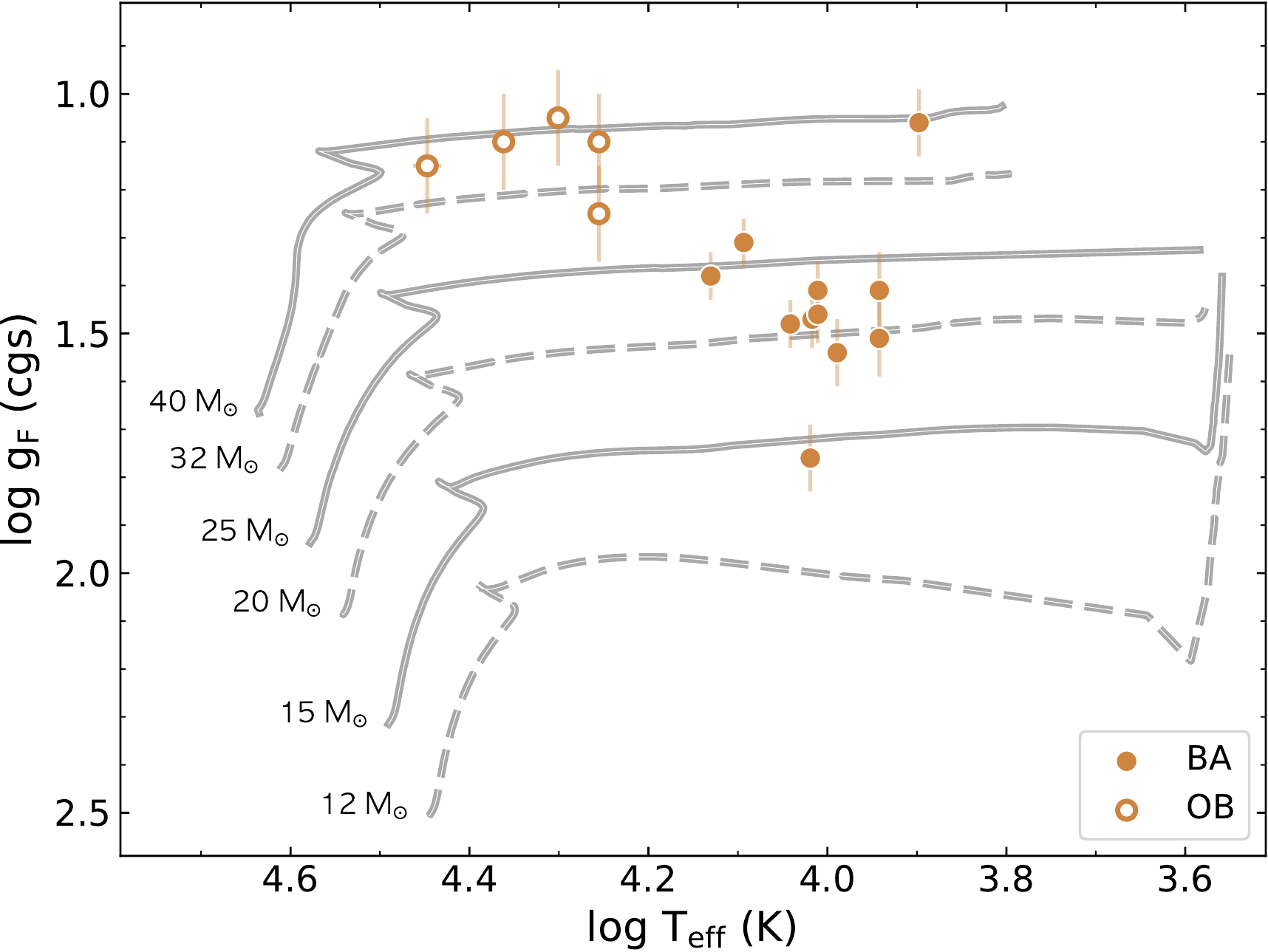}\medskip
	\caption{Spectroscopic Hertzsprung-Russell diagram for the blue supergiant sample of Table~\ref{table:2}. A selection of intermediate-mass evolutionary tracks with rotation from \citet{Ekstrom:2012} is included.}\label{fig:HRD}
\end{figure}

The location of the \bsg s in the spectroscopic Hertzsprung-Russell diagram (reporting the distance-independent \loggf\ \vs\ log\,\teff\ -- see \citealt{Langer:2014}) is displayed in Fig.~\ref{fig:HRD}. Stellar tracks that allow for the effects of rotation, taken from \citet{Ekstrom:2012}, are included in the diagram as a reference. From these evolutionary tracks we can infer stellar masses of our sample stars in the approximate range 15-40~\msun, in line with findings from our previous investigations in nearby galaxies (\eg\ \citealt{Kudritzki:2016, Bresolin:2016, Urbaneja:2017}).

\begin{figure}
	\center \epsscale{1.15}\plotone{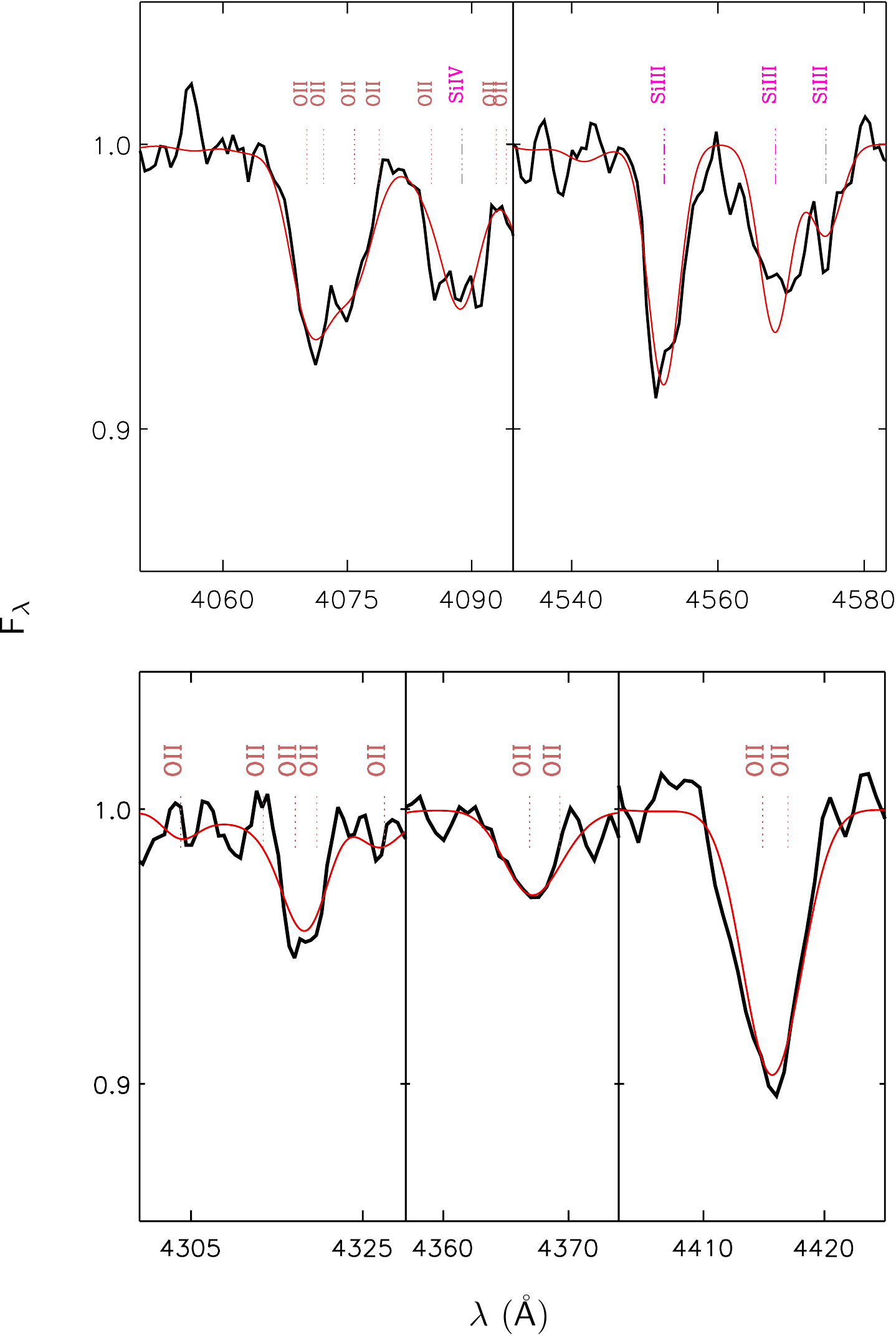}\medskip
	\caption{Fits of the model (red curve) to the observed (black curve) metal lines of star \target{43} (type B1~Ia) inside five wavelength windows. Lines from the main atomic species used to calculate the synthetic spectra are identified.}\label{fig:MetalFitB}
\end{figure}

\floattable
\begin{deluxetable}{lcccccccc}
	\tabletypesize{\footnotesize}	
	\tablecolumns{9}
	\tablewidth{0pt}
	\tablecaption{Stellar parameters.\label{table:2}}
	
	\tablehead{
		\colhead{ID}	     		&
		\colhead{\teff}	 			&
		\colhead{\logg}	 			&
		\colhead{\loggf}			&
		\colhead{[$Z$]}				&
		\colhead{\mbol}				&
		\colhead{\ebv}				&
		\colhead{BC$_{\rm V}$}				&		
		\colhead{Slit}\\[-2.5ex]
		\colhead{}	     		&
		\colhead{}	 			&
		\colhead{}	 			&
		\colhead{}				&
		\colhead{}				&
		\colhead{}				&
		\colhead{\scriptsize or}			&
		\colhead{\scriptsize or}				&		
		\colhead{}\\[-2.5ex]		
		\colhead{}	     		&
		\colhead{}	 			&
		\colhead{}	 			&
		\colhead{}				&
		\colhead{}				&
		\colhead{}				&
		\colhead{\textit{E(u\,$-$\,g)}}			&
		\colhead{\textit{BC$_g$}}				&		
		\colhead{}\\[-1ex]		
		\colhead{}       			&
		\colhead{(K)}       		&
		\colhead{(cgs)}       		&
		\colhead{(cgs)}				& 	
		\colhead{(dex)}				& 
		\colhead{(mag)}				& 	
		\colhead{(mag)} 	 		& 																			
		\colhead{(mag)}			 	& 																			
		\colhead{}  } 	
	\colnumbers			
	\startdata
	\\[-4.5ex]
	02   &   $10400\pm250$  &   $1.54\pm0.05$  &   $1.47\pm0.06$  &   $-0.37\pm0.10$  &   $19.56\pm0.14$  & $\mathit{0.15\pm0.03}$  & $\mathit{-0.27\pm0.05}$  &   B20 \\[-0.4ex] 
	03   &   $12400\pm200$  &   $1.69\pm0.05$  &   $1.31\pm0.05$  &   $-0.29\pm0.06$  &   $18.84\pm0.07$  & $\mathit{0.19\pm0.01}$  & $\mathit{-0.64\pm0.05}$ &   B19 \\[-0.4ex]
	05   &   $18000\pm500$  &   $2.27\pm0.10$  &   $1.25\pm0.10$  &   $-0.21\pm0.10$  &   $18.34\pm0.13$  & $\mathit{0.27\pm0.03}$  & $\mathit{-1.48\pm0.05}$  &   B21 \\[-0.4ex]
	08   &   $28000\pm1000$ &   $2.94\pm0.10$  &   $1.15\pm0.10$  &   $-0.24\pm0.10$  &   $17.35\pm0.10$  & $0.30\pm0.02$  & $-2.72\pm0.05$  &   B14 \\[-0.4ex]
	10   &   $11000\pm150$  &   $1.65\pm0.05$  &   $1.48\pm0.05$  &   $-0.15\pm0.10$  &   $19.52\pm0.04$  & $0.15\pm0.01$  & $-0.48\pm0.03$  &   B12 \\[-0.4ex] 
	16   &   $18000\pm500$  &   $2.12\pm0.10$  &   $1.10\pm0.10$  &   \m$0.02\pm0.10$ &   $17.06\pm0.13$  & $0.38\pm0.03$  & $-1.65\pm0.05$  &   A28 \\[-0.4ex] 
	18   &   $7900\pm30$  	&   $0.65\pm0.05$  &   $1.06\pm0.07$  &   $-0.22\pm0.10$  &   $17.45\pm0.11$  & $\mathit{0.25\pm0.02}$  & \m$\mathit{0.48\pm0.05}$  &   A27 \\[-0.4ex]
	20   &   $10250\pm250$  &   $1.45\pm0.05$  &   $1.41\pm0.06$  &   $-0.10\pm0.10$  &   $18.88\pm0.06$  & $0.20\pm0.01$  & $-0.34\pm0.05$  &   A31 \\[-0.4ex] 
	23   &   $20000\pm500$  &   $2.25\pm0.10$  &   $1.05\pm0.10$  &   $-0.15\pm0.10$  &   $17.76\pm0.10$  & $0.15\pm0.02$  & $-1.97\pm0.05$&   A30 \\[-0.4ex]
	25   &   $8750\pm100$  	&   $1.17\pm0.05$  &   $1.41\pm0.08$  &   \m$0.00\pm0.10$ &   $19.23\pm0.05$  & $0.19\pm0.01$  & $-0.05\pm0.03$  &   A24 \\[-0.4ex]
	28   &   $8750\pm100$  	&   $1.27\pm0.05$  &   $1.51\pm0.08$  &   $-0.15\pm0.10$  &   $19.88\pm0.05$  & $0.14\pm0.01$  & $-0.05\pm0.02$ &   B02 \\[-0.4ex] 
	29   &   $10250\pm300$  &   $1.50\pm0.05$  &   $1.46\pm0.06$  &   $-0.05\pm0.15$  &   $19.05\pm0.07$  & $0.16\pm0.01$  & $-0.33\pm0.06$ &   A19 \\[-0.4ex] 
	38   &   $9750\pm250$  	&   $1.47\pm0.05$  &   $1.54\pm0.07$  &   $-0.22\pm0.10$  &   $20.23\pm0.14$  & $\mathit{0.13\pm0.03}$  & $\mathit{-0.14\pm0.05}$  &   A14 \\[-0.4ex] 
	41   &   $13500\pm250$  &   $1.90\pm0.05$  &   $1.38\pm0.05$  &   $-0.33\pm0.10$  &   $18.95\pm0.07$  & $\mathit{0.23\pm0.02}$  & $\mathit{-0.82\pm0.03}$  &   A15 \\[-0.4ex] 
	43   &   $23000\pm500$  &   $2.55\pm0.10$  &   $1.10\pm0.10$  &   $-0.25\pm0.10$  &   $17.03\pm0.13$  & $\mathit{0.23\pm0.03}$  & $\mathit{-2.10\pm0.05}$  &   A09 \\[-0.4ex] 
	47   &   $10450\pm300$  &   $1.84\pm0.05$  &   $1.76\pm0.07$  &   $-0.35\pm0.10$  &   $19.52\pm0.17$  & $\mathit{0.29\pm0.04}$  & $\mathit{-0.26\pm0.06}$  &   A01 \\[-0.4ex] 
	\\[-2.5ex]
	\enddata
	\tablecomments{Columns 7 and 8 report the reddening and bolometric correction, respectively, based either on 
		the HST (non italicized) or the CFHT (italicized) photometry.}
\end{deluxetable}

\section{Metallicities: stars \vs\ ionized gas} \label{sec:comparison}
One of our primary aims in carrying out the spectroscopy of extragalactic \bsg s is to evaluate the present-day
chemical abundance of galaxies, independent of, and complementary to, other methods, in particular the study of the emission lines of \hii\ regions. This bears on the study of both radial abundance gradients in spiral galaxies and the galaxy mass-metallicity relation.


In order to compare nebular and stellar chemical abundances in NGC~2403 we examine the radial abundance gradients obtained independently from \hii\ regions and \bsg s. The main caveat in this approach is that the nebular abundances typically refer only to oxygen, while, as explained in Sect.~\ref{sec:quantitative}, the stellar abundances are calculated using stellar features originating from several chemical elements. We assume solar metal abundance ratios and express the metallicity of the ionized gas by adopting the solar oxygen abundance value, \eosun\,=\,\ohsun\,8.69 (\citealt{Allende-Prieto:2001}).

The most comprehensive study of the \hii\ region chemical abundances in NGC~2403 todate is the work published by \citet{Rogers:2021}. These authors used direct (\ie\ based on the detection of auroral lines) abundances of 27 nebulae to calculate the exponential oxygen abundance gradient in this galaxy. In order to facilitate the comparison with their results, we adopt the same deprojection parameters for the computation of the galactocentric distances of our targets (see footnote to Table~\ref{table:1}). 
\citet{Rogers:2021} carried out a regression (linear since dealing with logarithmic abundances)
accounting for uncertainties in both \eo\ and the galactocentric distances, assuming a 5\% error in $r/$\re. We follow the same approach, and adopt the {\sc linmix}\footnote{\url{https://github.com/jmeyers314/linmix}} package, a Python port of the Bayesian method described by \citet{Kelly:2007}, for calculating the linear regression. Finally, we perform the analysis using galactocentric distances normalized to the galactic effective radius \re, as done by \citet{Rogers:2021}.

\begin{figure*}
	\center\includegraphics[width=0.8\textwidth]{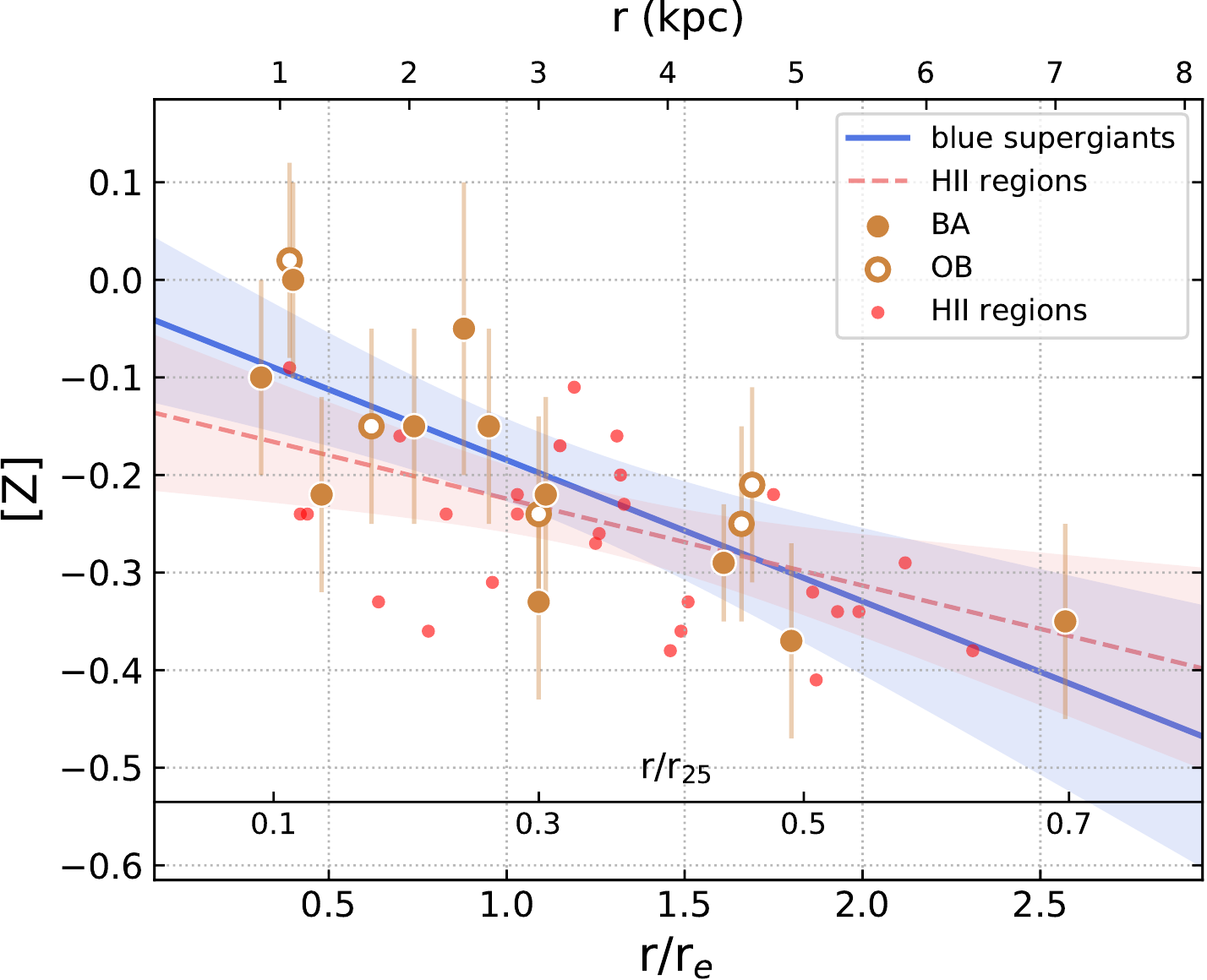}\medskip
	\caption{The galactocentric metallicity radial gradient of NGC~2403 from both blue supergiants and \hii\ regions. {\em Blue continuous line:} regression determined using the 16 blue supergiants of Table~\ref{table:2}, represented by closed (BA) and open (OB) circles. {\em Red dashed line:} regression determined using the oxygen abundances of the \hii\ regions (\citealt{Rogers:2021}), represented by small red dots. For the sake of clarity we omit the errorbars of the \hii\ region data points. The shaded areas represent the 95\% confidence intervals of the fits.}\label{fig:Gradient}
\end{figure*}

In Fig.~\ref{fig:Gradient} we display the stellar metallicity gradient, determined from 16 data points, using circle symbols for the \bsg\ measurements (Sect.~\ref{sec:quantitative}, and Table~\ref{table:2}). The equation of the linear regression, shown by the continuous blue line, is:

\begin{equation}
[Z] = -0.04\, (\pm 0.06) - 0.14\, (\pm 0.05)~ r/r_e
\end{equation}

\noindent
while for the \hii\ regions (dashed red line -- 27 data points) we obtain

\begin{equation}
[Z] = -0.14\, (\pm 0.04) - 0.09\, (\pm 0.03)~ r/r_e
\end{equation}

\noindent
which replicates Eq.\,(9) of \citet{Rogers:2021}, once \eo\ is expressed in terms of metallicity [Z].
The intrinsic scatter about the regression lines for \bsg s ($\sigma_i = 0.05\pm0.03$ dex) and ionized nebulae ($\sigma_i = 0.04\pm0.02$ dex) is comparable, as is the standard error of the regression (0.07 dex). 

We tested the agreement between the regressions derived for \bsg s and \hii\ regions 
with a bootstrap approach (\citealt{Efron:1979}). We found that the differences in the slope and intercept values between the two regressions both fall inside the corresponding 95\% confidence intervals.
In order to test the null hyphoteses that slopes and intercepts are equal, we considered the studentized, bootstrap-$t$ interval 
(\citealt{Efron:1994}; see example in \citealt{Wehrens:2000}) and found $p$-values $p$\,=\,0.13 (slopes) and $p$\,=\,0.08 (intercepts). Adopting the typical critical value $\alpha=0.05$, we conclude that both the slopes and the intercepts of the two regressions are not significantly different.

We should add, however, that the conclusion concerning the intercepts should be considered with some caution, since it depends on both the solar \eosun\ value we adopt and, arguably more importantly, on how oxygen depletion on dust grains affects the nebular abundances. A correction to the nebular metallicities of approximately +0.1 dex should be expected from empirical results in the literature 
(\citealt{Mesa-Delgado:2009, Peimbert:2010}).

Utilizing the bootstrap method we also verified  that the regressions to the OB and BA supergiants, taken separately, can be considered statistical equivalent. This result lends further support to our combining these two groups of stars, which are analyzed with different techniques and with different metal line diagnostics (Sect.~\ref{sec:quantitative}).

If we merge the stellar and nebular data sets, we obtain the linear gradient

\begin{equation}
[Z] = -0.11\, (\pm 0.03) - 0.11\, (\pm 0.03)~ r/r_e.
\end{equation}

For completeness, we express the slope of the stellar abundance gradient in units of dex\,\,\rtf$^{-1}$,  dex\,\,kpc$^{-1}$ and dex\,\,\rd$^{-1}$ (\rd\ is the disk scale length, with \re\,=\,1.678~\rd) as follows:

\begin{equation}
\nabla_{r_{25}} =  -0.53 \pm 0.18
\end{equation}

\begin{equation}
\nabla_{kpc} =  -0.05 \pm 0.02
\end{equation}

\begin{equation}
\nabla_{r_{d}} =  -0.09 \pm 0.03
\end{equation}

\subsection{Stellar and gaseous metallicities: review of literature data}
In \citet{Bresolin:2016} we presented a comparison between \bsg\ and ionized gas metallicities obtained for 14 galaxies (the Orion nebula representing the Milky Way). The nebular chemical abundances were obtained utilizing the direct method, which is based on the fluxes of auroral lines and of stronger collisionally excited lines (\cel s). For seven systems the nebular abundances were evaluated also from the strengths of \oiiION\ recombination lines (\rl s), whose analysis typically leads to $\sim$0.2 dex higher abundances compared to the \cel\ analysis (the abundance discrepancy factor ADF, \citealt{Garcia-Rojas:2007}). 

\begin{figure*}
	\center\includegraphics[width=0.8\textwidth]{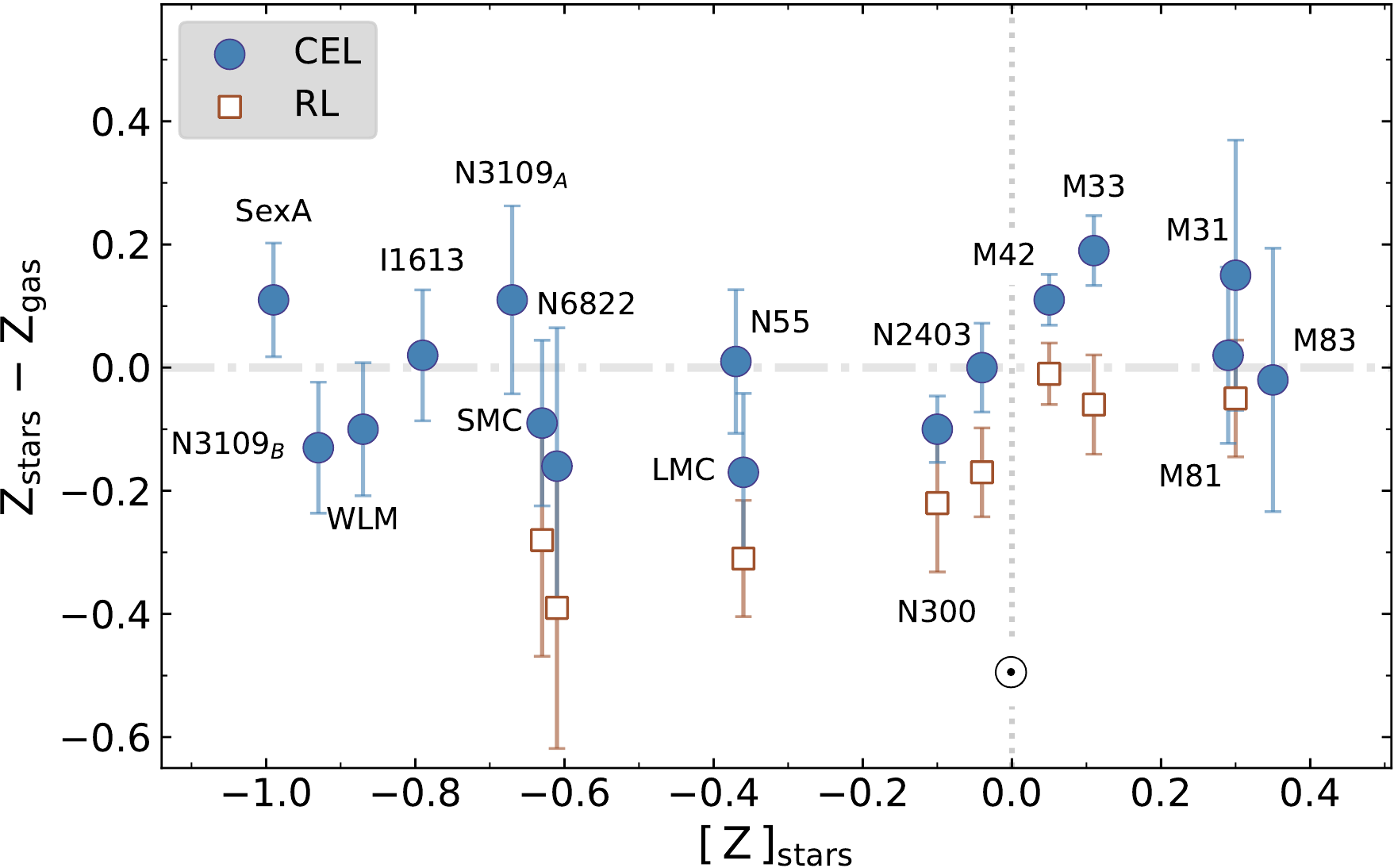}\medskip
	\caption{Difference in the metallicities derived from blue supergiants and \hii\ regions as a function of stellar metallicity
		for a galaxy sample drawn from the literature (after \citealt{Bresolin:2016}) and NGC~2403. The nebular metallicities are increased by 0.1 dex, in order to account for oxygen depletion onto dust grains. The \hii\ region chemical abundances are obtained either from collisionally excited lines (\cel s) with the direct method (blue dots) or from \opp\ recombination lines (\rl s, red squares).\\}\label{fig:stars_hii_RL}
\end{figure*}

We update  the comparison presented by \citet{Bresolin:2016} by including the results obtained here for NGC~2403,
as well as taking into account more recent stellar metallicities for \bsg s in M31 and M33, recently studied by \citet{Liu:2022}. The outcome is shown in Fig.~\ref{fig:stars_hii_RL}, where we follow our original procedure: (a) for irregular galaxies we adopt mean abundances; (b) for spirals we adopt the central abundances inferred from the abundance gradients (except for the Milky Way, since we only use M42 as representative of the solar neighborhood); (c) we add 0.1 dex to the nebular metallicities in order to approximately correct for the depletion of oxygen onto dust grains.

The addition of NGC~2403 to the diagram in Fig.~\ref{fig:stars_hii_RL} is based on the stellar metallicities we present here, the \hii\ region \cel\ abundances of \citet{Rogers:2021} and the \rl\ detection of \citet{Esteban:2009} in a single \hii\ region (VS44 in their paper, equivalent to NGC~2403+96+30 in \citealt{Rogers:2021}). \citet{Esteban:2009} and \citet{Rogers:2021} report 
the same \opp\ abundance for this nebula, thus we apply ADF(\opp)\,=\,0.30 from \citet{Esteban:2009} to augment the 
\citet{Rogers:2021} total O/H abundance by 0.17 dex. We finally apply this correction to the \hii\ region regression intercept in order to obtain the NGC~2403 \rl\ data point in Fig.~\ref{fig:stars_hii_RL}.

Looking at Fig.~\ref{fig:stars_hii_RL} we cannot but confirm the conclusions drawn by \citet{Bresolin:2016}: on average the stellar and the direct nebular metallicities are in good agreement across the whole $[Z]$ range, even though at high metallicities the former tend to be ~$\sim$~0.1-0.2 dex higher for the two objects with the smallest error bars (M42 and M33). In the high-metallicity regime the stellar metallicities are in agreement with the \rl-based nebular abundances. On the other hand, we observe a  discrepancy of the \rl-based metallicities with decreasing $[Z]$ relative to the stellar metallicities. The new data points that refer to NGC~2403 are in line with our initial interpretation. We should add that the apparent trend seen for the \rl-based nebular metallicities could simply be reflecting a similar trend for the \cel-based metallicities, since the ADF is roughly constant. Attempts by the first author to detect \oiiION\ \rl s at the low-metallicity end (WLM, IC~1613) have proved unsuccessful so far. However, we remark that below the solar value the direct nebular metallicities are in good agreement with the \bsg s across approximately one order of magnitude in $[Z]$, while the  nebular metallicities based on \rl s are not. For further discussion we refer to 
\citet{Bresolin:2016}.

\section{The stellar mass-metallicity relation} \label{sec:mzr}
In our previous work (\citealt{Kudritzki:2012, Kudritzki:2016, Hosek:2014, Bresolin:2016, Davies:2017})
we presented the stellar mass-metallicity relation (\mzr) of local galaxies based on stellar abundance determinations, in place of the more commonly adopted nebular abundances. Following the same spirit that motivated the comparison carried out in Sect.~\ref{sec:comparison}, it is critical to validate the results found through the emission line analysis of extragalactic \hii\ regions with independent determinations of galaxy metallicities secured via the absorption line analysis of the stellar component. 

\begin{figure}
	\epsscale{1.15}\center\plotone{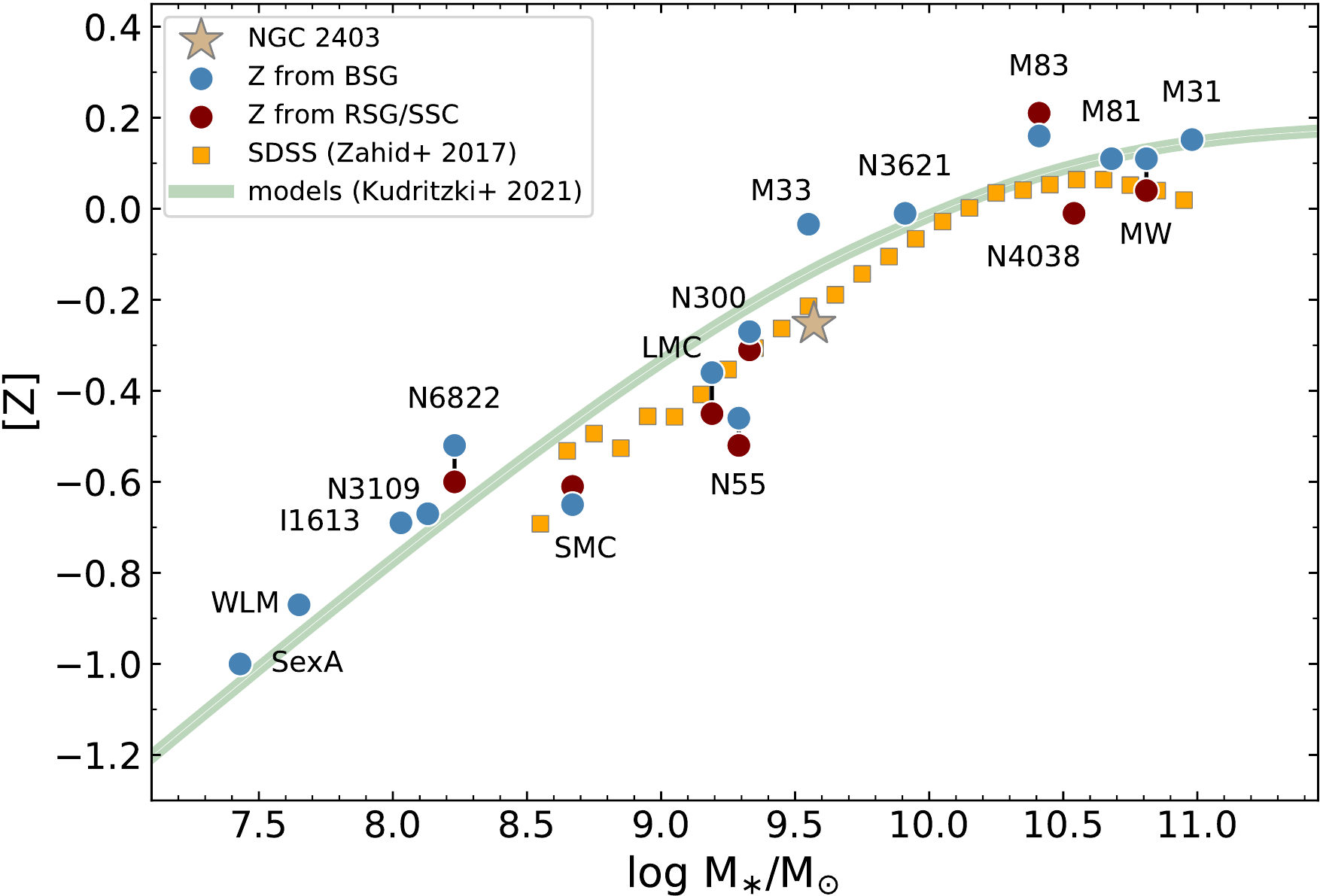}\medskip
	\caption{The stellar mass-metallicity relation of local galaxies based on abundance data derived from the analysis of stellar spectral features: blue supergiants (\bsg), red supergiants (\rsg), super star clusters (\ssc) and SDSS galaxies. Predictions from the look-back models of \citet{Kudritzki:2021} are included for comparison (green curve).}\label{fig:mzr}
\end{figure}

In Fig.~\ref{fig:mzr} we update the stellar \mzr\ with our new NGC~2403 \bsg\ data, as well as with the 
results of the study of M31 and M33 by \citet{Liu:2022}. Because of the presence of abundance gradients in spiral galaxies, we, somewhat arbitrarily, select the metallicity at a galactocentric distance of 0.4\,\rtf\ as representative for the whole galaxies. For NGC~2403 we adopt the stellar mass value \logm\,=\,9.57, from \citet{Leroy:2019}. The location of NGC~2403 in our diagram (stellar symbol) is in line with the trend outlined by the additional galaxies where \bsg s have already been investigated (blue dots).

Fig.~\ref{fig:mzr} includes two additional sets of independent data points. First, those proceeding from the chemical abundance analysis of near-IR spectra of individual red supergiants (\rsg s) and super star clusters (\ssc s), whose light output is dominated by \rsg s (red dots). Both the \rsg s and the \ssc s are representative of the young population in galaxies, similarly to the \bsg s and the \hii\ regions, and can therefore be displayed together in the diagram. In Appendix~A we provide the details of the stellar abundances and masses used to produce Fig.~\ref{fig:mzr}, together with the references to the original abundance work. 
In addition, in the figure we display the empirical relation obtained by \citet[yellow squares]{Zahid:2017} from the population synthesis of the stacked SDSS spectra of $2\times10^5$ local star-forming galaxies ($0.027 < z < 0.25$). The \mzr\ delineated by the SDSS galaxies is in accordance with the relation defined by the supergiant and \ssc\ data.

Finally, the green curve in Fig.~\ref{fig:mzr} represents the \mzr\ resulting from the chemical look-back evolution models by \citet{Kudritzki:2021}. Despite not being a fit to the observed data, the model curve 
is compatible with the empirical \mzr\ we observe from stellar metallicities.

\section{The spectroscopic distance to NGC~2403} \label{sec:distance}

Having obtained the stellar parameters as described in Sect.~\ref{sec:quantitative}, we can measure the spectroscopic distance to NGC~2403 via the flux-weighted gravity--luminosity relationship (\fglr; \citealt{Kudritzki:2003, Kudritzki:2008}) and the stellar mass-luminosity relationship.
The existence of a correlation between the flux-weighted gravity \gf\,$\equiv$\,$g$/\tefffour\ and the bolometric magnitude \Mbol\  stems from the virtually constant luminosities and masses of \bsg s as they evolve. 

We adopt the calibration of the \fglr\ by \citet{Urbaneja:2017}, established from the analysis of 90 \bsg s in the Large Magellanic Cloud, but reformulated in order to account for the slightly smaller distance modulus to this galaxy reported by \citet{Pietrzynski:2019}, $(m-M)_{LMC} = 18.477 \pm 0.004\ {\rm (statistical)} \pm 0.026\ {\rm (systematic)}$. The relationship presented by \citet{Urbaneja:2017} is a piecewise linear fit to the stellar parameters, characterized by a single break point. 
With the modified distance to the LMC, the \fglr\ is:

\begin{equation}\label{eq:fglr1}
M_{bol} = 3.20\, (\log\,g_{\mbox{\tiny\it F}} - 1.5) - 7.878
\end{equation}

\noindent
for \loggf\,$>$\,1.30, and 

\begin{equation}\label{eq:fglr2}
M_{bol} = 8.34\, (\log\,g_{\mbox{\tiny\it F}} - 1.3) - 8.518
\end{equation}

\noindent
for \loggf\,$<$\,1.30.

\begin{figure}
	\epsscale{1.15}\center\plotone{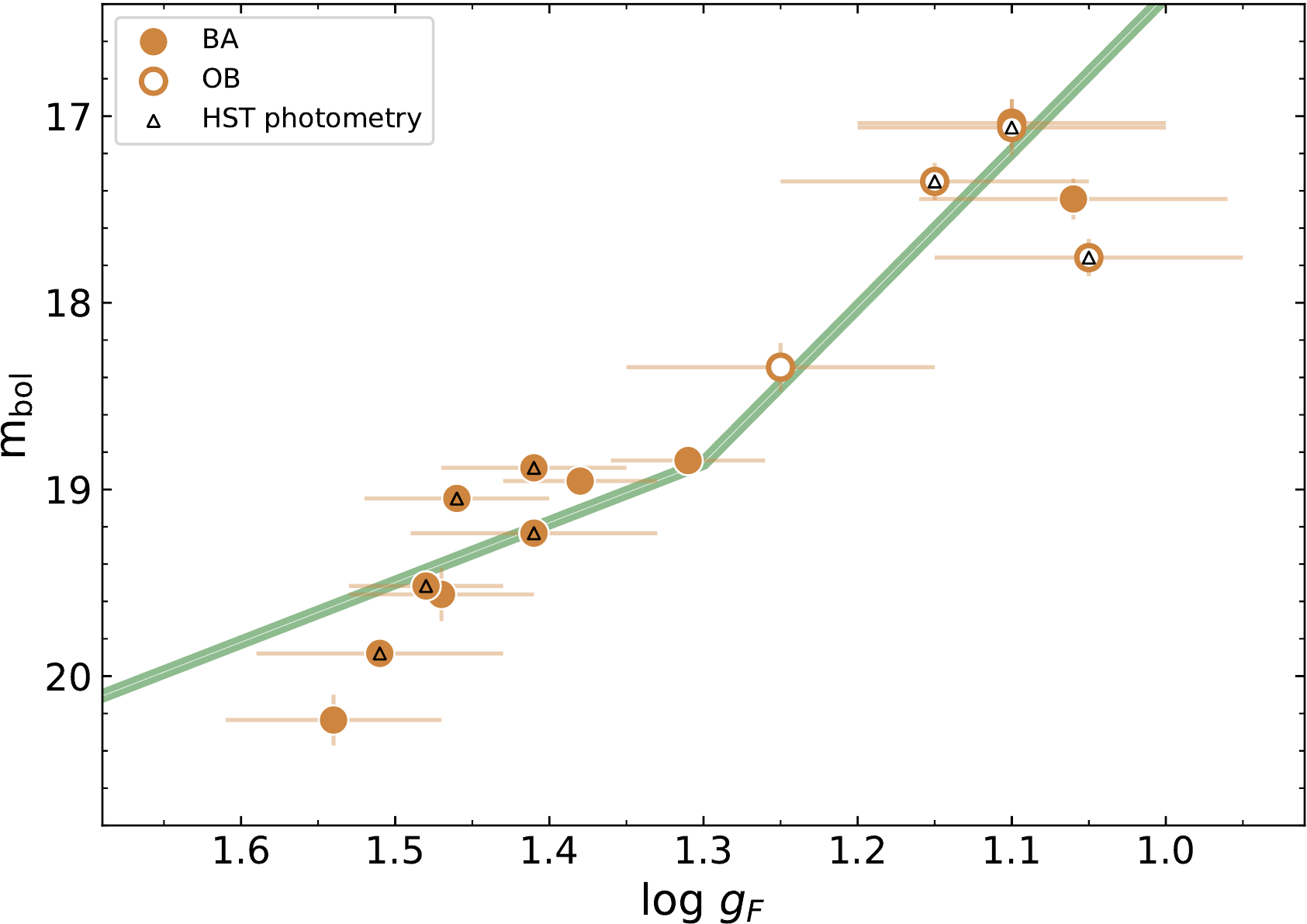}\medskip
	\caption{The \fglr\ in NGC~2403. The data points (stars from Table~\ref{table:2}, except for \target{47})
	are fitted with the fiducial relation (Eq.~\ref{eq:fglr1}-\ref{eq:fglr2}), vertically shifted by an amount equivalent to the distance modulus. The fitting relation is represented by the green line. }\label{fig:fglr}
\end{figure}

The empirical \fglr\ traced by the NGC~2403 \bsg s is illustrated in Fig.~\ref{fig:fglr}, where
we excluded \target{47} since it is a bright outlier, most likely a blend. There are no obvious differences in the distribution of the data points, whether the source of the photometry is either HST (markers with inner triangles) or CFHT (no triangles), so we believe that in general blending is not a major issue for the ground-based data. 

The distance modulus can be assessed by the vertical shift of the fiducial relation (Eq.~\ref{eq:fglr1} and \ref{eq:fglr2}) that produces the best fit (green line) to the observed distribution of points in Fig.~\ref{fig:fglr}.
An orthogonal distance regression fit, accounting for errors in both coordinates and computed with the {\tt scipy.odr} Python package, yields

$$\mu = m-M = 27.38 \pm 0.08$$ \noindent which corresponds to a distance 
of $2.99^{+0.10}_{-0.12}$~Mpc. 

\subsection{Comparison to other stellar distance indicators}
NGC~2403 was among the first galaxies outside the Local Group where Cepheid variables were identified and studied by Hubble and collaborators (\citealt{Tammann:1968}).
The period-luminosity relation of 10 Cepheids in NGC~2403 has been investigated from ground-based CCD observations by \citet{Freedman:1988}. The distance modulus they derived, $\mu = 27.51 \pm 0.24$, was revised by \citet{Freedman:2001} to $\mu = 27.48 \pm 0.10$. We report the result they obtained without correction for metallicity effects, because the metallicity relative to the LMC that \citet{Freedman:2001} adopt, $\Delta Z = 0.3$ dex, is quite large and not consistent with the measurements we show in Fig.~\ref{fig:stars_hii_RL}. 
It should be noted that only $I$-band data were used, and that a reddening value $E(V-I)=0.20\pm0.10$ was adopted, not measured. From our sample in Table~\ref{table:2} we transformed the reddening values to $E(V-I)$ using our models, and determined an average $E(V-I)=0.27\pm0.09$. This higher reddening would decrease the \citet{Freedman:2001} distance modulus by 0.1 magnitudes. In addition, the zero point of the Cepheid period-luminosity relation used by these authors is defined by
the distance modulus of the LMC, which they adopted to be 18.5. The \fglr\ distance is instead tied to a $\sim$0.02 mag smaller value from \citet{Pietrzynski:2019}. Overall, these effects would lead to an excellent agreement between a `corrected' Cepheid distance ($\mu = 27.36\pm0.10$) and our \fglr\ determination.

In addition to the $I$-band data, \citet{Saha:2006} included $B$-band photometry from \citet{Tammann:1968} and obtained
$\mu = 27.44 \pm 0.15$. The definition of their zero point contains a distance modulus of the LMC of 18.54, which is 0.06 mag larger than the value we adopt. 
Our \fglr\ measurement is again fully consistent with their Cepheid distance.

\begin{figure}
	\epsscale{1.15}\center\plotone{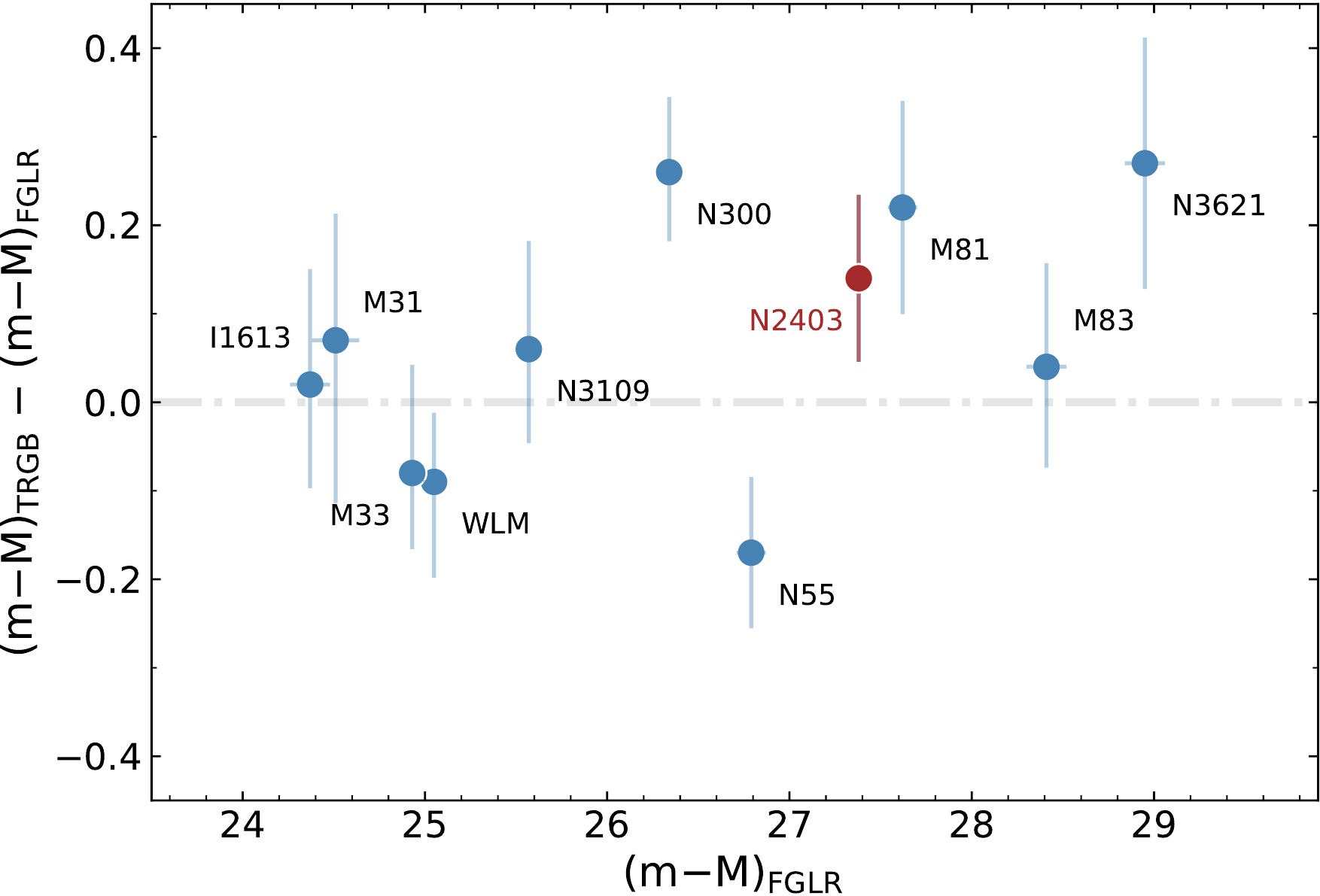}\medskip
	\caption{Distance modulus offset between \trgb\ and \fglr\ determinations as a function of the \fglr\ distance modulus for 11 galaxies.}\label{fig:fglr_vs_trgb}
\end{figure}

More accurate distance determinations have been published using the F814W magnitude of the Tip of the Red Giant Branch (\trgb) measured in various HST fields of NGC~2403. A weighted mean of the distance moduli determined from one WFPC2 and two ACS fields by \citet{Dalcanton:2009} yields $\mu = 27.51 \pm 0.03$. 
Similarly, from three ACS fields analyzed by \citet{Radburn-Smith:2011} we obtain 
$\mu = 27.51 \pm 0.05$. The analysis presented by the Extragalactic Distance Database (EDD: \citealt{Tully:2009}) yields $\mu = 27.52 \pm 0.05$ using F814W magnitudes, as well as values of $27.56\pm0.02$ (F110W) and $27.53\pm0.01$ (F160W).
Therefore, there is an offset $\Delta\mu = \mu_{TRGB} - \mu_{FGLR}$ relative to our \fglr\ result, which is highly significant if we consider only the near-IR \trgb\ work.
In Fig.~\ref{fig:fglr_vs_trgb} we follow \citet{Sextl:2021} and plot $\Delta\mu$, adopting the EDD F814W-based values for the \trgb\ distances\footnote{\citet{Sextl:2021} use the mean of the EED and ANGST values for four of the galaxies, while we only use the EDD.}, as a function of the \fglr\ distances we have determined in our previous work on 10 galaxies\footnote{The published \fglr\ distances have been adjusted, when needed, to account for the \citet{Urbaneja:2017} calibration and the  \citet{Pietrzynski:2019} LMC distance.}, as well as for NGC~2403. Table~\ref{table:FGLR_TRGB}
summarizes the distance moduli we adopt.

The offset we find in the case of NGC~2403, $\Delta\mu = 0.14\pm0.09$, is significant at the ~$\sim$1.5$\sigma$ level.
We find comparable or larger offsets for other galaxies (NGC~300, M81, NGC~3621), and the mean difference is $\overline{\Delta\mu} = 0.067 \pm\ 0.044$. The data points in Fig.~\ref{fig:fglr_vs_trgb} may visually suggest a potential trend with distance, but a correlation analysis based on 
both the Pearson and Spearman correlation coefficients rules this out (Pearson's $r=0.47$; $p$-value for the null hypothesis that there is no correlation $p=0.14$).

\begin{deluxetable}{lcc}
	\tabletypesize{\footnotesize}	
	\tablecolumns{3}
	\tablewidth{0pt}
	\tablecaption{\fglr\ and \trgb\ distance moduli.\label{table:FGLR_TRGB}}
	
	\tablehead{
		\colhead{Galaxy}	     		&
        \multicolumn{2}{c}{distance modulus}	\\[-2.5ex]	
		%
		%
		\colhead{}       			&
		\colhead{\fglr}       		&
		\colhead{\trgb}   }    		
	\startdata
		\\[-10pt]
	IC 1613    	&   24.37 $\pm$ 0.11\tablenotemark{\scriptsize a}   &   24.39 $^{+0.07}_{-0.04}$   	\\[4pt] 
	M31			&	24.51 $\pm$ 0.13\tablenotemark{\scriptsize b}   &   24.58 $^{+0.06}_{-0.13}$   	\\[4pt] 
	M33        	&   24.93 $\pm$ 0.07\tablenotemark{\scriptsize b}   &   24.85 $^{+0.10}_{-0.05}$  	\\[4pt]
	WLM			&	25.05 $\pm$	0.06\tablenotemark{\scriptsize c}	&	24.96 $^{+0.05}_{-0.09}$	\\[4pt]
	NGC 3109	&	25.57 $\pm$	0.07\tablenotemark{\scriptsize d}	&	25.63 $^{+0.10}_{-0.08}$	\\[4pt]
	NGC 300		&	26.34 $\pm$	0.06\tablenotemark{\scriptsize e}	&	26.60 $^{+0.06}_{-0.05}$	\\[4pt]
	NGC 55		&	26.79 $\pm$	0.08\tablenotemark{\scriptsize f}	&	26.62 $^{+0.03}_{-0.03}$	\\[4pt]
	NGC 2403	&	27.38 $\pm$	0.08\tablenotemark{\scriptsize g}	&	27.52 $^{+0.05}_{-0.05}$	\\[4pt]
	M81			&	27.62 $\pm$	0.08\tablenotemark{\scriptsize h}	&	27.84 $^{+0.09}_{-0.09}$	\\[4pt]
	M83			&	28.41 $\pm$	0.11\tablenotemark{\scriptsize i}	&	28.45 $^{+0.04}_{-0.03}$	\\[4pt]
	NGC 3621	&	28.95 $\pm$	0.11\tablenotemark{\scriptsize j}	&	29.22 $^{+0.09}_{-0.09}$	\\
	\\[-2.5ex]
	\enddata
\tablecomments{The references to the original \fglr\ data are given below. The \fglr\ distances are homogenized to the
same calibration and zero point. The \trgb\ distances are extracted from the Extragalactic Distance Database (\citealt{Tully:2009}).	}
\tablerefs{$^a$\citet{Berger:2018}; $^b$\citet{Liu:2022}; $^c$\citet{Urbaneja:2008}; $^d$\citet{Hosek:2014};
$^e$\citet{Kudritzki:2008}; $^f$\citet{Kudritzki:2016}; $^g$This paper; $^h$\citet{Kudritzki:2012}; 
$^i$\citet{Bresolin:2016}; $^j$\citet{Kudritzki:2014}.}	
\end{deluxetable}

\section{Summary} \label{sec:summary}
We have carried out the first quantitative spectral analysis of blue supergiant stars in the nearby galaxy NGC~2403. Out of a sample of 47 targets observed with the LRIS spectrograph at the Keck~I telescope we have extracted 16 BA stars for which we have spectra of sufficient quality to carry out a comparison with model spectra of evolved massive stars in order to derive the stellar parameters. 
There are two main outcomes from our study:

\begin{enumerate}
\item We measured the galactocentric radial metallicity gradient of NGC~2403 based on our stellar observations, finding that it agrees well with the results obtained from \hii\ regions,  both in slope and intercept, especially if we account for a $\sim$0.1 dex depletion of the nebular oxygen due to dust grains. This lends further evidence that the extragalactic stellar metallicities we obtain from our technique are in general agreement with the nebular abundances based on the analysis of the auroral lines over more than one order of magnitude in metallicity.

\item Adopting the known relation between stellar parameters and intrinsic luminosity (\fglr) we derived a distance to NGC~2403 of $2.99^{+0.10}_{-0.12}$~Mpc ($\mu = 27.38 \pm 0.08$ mag). While this can be brought into agreement with Cepheid-based determinations, once we account for differences in reddening and LMC distance between the separate studies,
it is 0.14 mag short of the \trgb\ value. We have no current explanation for this discrepancy, but note a similar behavior in a handful of spiral galaxies.
\end{enumerate}

\begin{acknowledgments}
We are indebted to Zach Gazak for his contributions in the early phases of this project and to
Richard Gray for sharing the spectra from his digital spectral classification atlas.
This research has made use of the Keck Observatory Archive (KOA), which is operated by the W. M. Keck Observatory and the NASA Exoplanet Science Institute (NExScI), under contract with the National Aeronautics and Space Administration.
RPK acknowledges support by the Munich Excellence Cluster Origins funded by the Deutsche
Forschungsgemeinschaft (DFG, German Research Foundation) under Germany's Excellence Strategy EXC-2094 390783311.
\end{acknowledgments}

\facility{Keck:I (LRIS)}

	
\software{APLpy (\citealt{Robitaille:2012, Robitaille:2019}), LINMIX (Meyers 2015, \url{https://github.com/jmeyers314/linmix}), SciPy (\citealt{Virtanen:2020}), NumPy (\citealt{Harris:2020}), Matplotlib (\citealt{Hunter:2007a}), PyRAF (\citealt{Science-Software-Branch-at-STScI:2012}).}

\appendix

\section{Values and sources of the stellar mass-metallicity relation}
We list here the galaxy stellar masses and metallicities that are used to construct the stellar mass-metallicity relation shown in Fig.~\ref{fig:mzr}. As explained in the text, the stellar metallicity data refer to blue supergiants (\bsg), red supergiants (\rsg) and super star clusters (\ssc). In Table~\ref{table:MZR} we divide our \logm\ and [Z] values into these three categories. The table also reports the literature sources for the measurements.\\

\floattable
\begin{deluxetable}{lcccc}
	\tabletypesize{\footnotesize}	
	\tablecolumns{5}
	\tablewidth{0pt}
	\tablecaption{Stellar mass-metallicity relation.\label{table:MZR}}
	
	\tablehead{
		\colhead{Galaxy}	     		&
		\colhead{\logm}	 			&
		\colhead{\m[$Z$]}	 			&
        \multicolumn{2}{c}{source}	\\[-2.5ex]	
		%
		%
		\colhead{}       			&
		\colhead{}       		&
		\colhead{}       		&
		\colhead{mass}				& 													
		\colhead{metallicity}  } 	
	\startdata
\cutinhead{BSG}
	M31        &   10.98  &   \m0.07     & \citet{Chemin:2009}    & \citet{Liu:2022}    \\ 	
	M81        &   10.68  &   \m0.11     & \citet{Leroy:2019}    & \citet{Kudritzki:2012}  \\
	Milky Way  &   10.81  &   \m0.11     & \citet{Sofue:2009}    & \citet{Przybilla:2008}    \\
	M83        &   10.41  &   \m0.16     & \citet{Leroy:2019}    & \citet{Bresolin:2016}    \\	
	NGC 3621   &   9.91   &   $-$0.01  & \citet{Leroy:2019}    & \citet{Kudritzki:2014} \\
	NGC 2403   &   9.57   &   $-$0.25  & \citet{Leroy:2019}    &  this work   \\	
	M33        &   9.55   &   $-$0.20  & \citet{Woo:2008}    & \citet{Liu:2022}    \\
	NGC 55     &   9.29   &   $-$0.46  & \citet{Kudritzki:2016}   & \citet{Kudritzki:2016} \\
	LMC        &   9.19   &   $-$0.36  & \citet{Woo:2008}    & \citet{Hunter:2007, Urbaneja:2017} \\ 
	NGC 300    &   9.33   &   $-$0.27  & \citet{Munoz-Mateos:2015}    & \citet{Kudritzki:2008} \\ 
	SMC        &   8.67   &   $-$0.65  & \citet{Woo:2008}   & \citet{Trundle:2005, Schiller:2010} \\ 
	NGC 6822   &   8.23   &   $-$0.52  & \citet{Woo:2008}    & \citet{Venn:2001} \\
	NGC 3109   &   8.13   &   $-$0.67  & \citet{Woo:2008}    & \citet{Hosek:2014} \\ 
	IC 1613    &   8.03   &   $-$0.69  & \citet{Woo:2008}    & \citet{Bresolin:2007a} \\ 
	WLM        &   7.65   &   $-$0.87  & \citet{Woo:2008}    & \citet{Urbaneja:2008} \\ 
	Sextans A  &   7.43   &   $-$1.00  & \citet{Woo:2008}    & \citet{Kaufer:2004} \\ 
\cutinhead{RSG}
    Milky Way  &   10.81  &   \m0.11     & \citet{Sofue:2009}    & \citet{Gazak:2014a}    \\ 
	NGC 55     &   9.29   &   $-$0.52  & \citet{Kudritzki:2016}   & \citet{Patrick:2017}  \\
    LMC        &   9.19   &   $-$0.45  & \citet{Woo:2008}    & \citet{Davies:2015} \\
	NGC 300    &   9.33   &   $-$0.31  & \citet{Munoz-Mateos:2015}    & \citet{Gazak:2015} \\
	SMC        &   8.67   &   $-$0.61  & \citet{Woo:2008}   & \citet{Davies:2015} \\  
	NGC 6822   &   8.23   &   $-$0.60  & \citet{Woo:2008}    & \citet{Patrick:2015} \\
\cutinhead{SSC}
    M83        &   10.41  &   \m0.21     & \citet{Leroy:2019}    & \citet{Davies:2017}    \\
    NGC 4038   &   10.54  &   $-$0.01     & \citet{Leroy:2019}    & \citet{Lardo:2015}    \\	
	\\[-2.5ex]
	\enddata
\tablecomments{(1) The [Z] values refer to galactocentric distances of 0.4\,\rtf\ in those cases where there is a chemical abundance gradient, \ie\ for the spiral galaxies. 
(2) In order to account for the different solar abundance patterns adopted in the BSG and RSG spectral analysis, the RSG metallicities are rescaled as explained in Appendix~A of \citet{Davies:2017}. 	
}	
\end{deluxetable}

\clearpage

\end{document}